%% file: starlink_edge.tex
\pdfoutput=1
\documentclass[10pt, conference, letterpaper]{IEEEtran}

\PassOptionsToPackage{hyphens}{url}\usepackage[colorlinks]{hyperref}

\usepackage{listings}
\usepackage{soul}
\usepackage{pgf}
\usepackage{multirow}
\usepackage{subfigure}
\usepackage[font=footnotesize,format=plain,labelformat=simple,labelsep=period]{caption}
\usepackage{balance}

\usepackage{cite}

\definecolor{dkgreen}{rgb}{0,0.6,0}
\definecolor{gray}{rgb}{0.5,0.5,0.5}
\definecolor{mauve}{rgb}{0.58,0,0.82}

\usepackage{letltxmacro}

\makeatletter
\def\blfootnote{\xdef\@thefnmark{}\@footnotetext}
\makeatother

\def\ver{2}

\def\StartVersionOne{\ifnum\ver=1}

\def\StartVersionTwo{\ifnum\ver=2}

\DeclareUnicodeCharacter{02B9}{'}

\newcommand{\formatfig}{\sffamily\tiny}
\newcommand{\formatnormal}{\normalsize\rmfamily}
\newcommand{\blind}[1]{\ifnum\ver=1 #1\fi}
\newcommand{\normal}[1]{\ifnum\ver=2 #1\fi}

\begin{document}

  \font\myfont= cmr11 at 21pt

\title{\huge Network Characteristics of LEO Satellite Constellations: \\A Starlink-Based Measurement from End Users}

\iftrue
\author{\IEEEauthorblockN{Sami Ma\IEEEauthorrefmark{1}, Yi Ching Chou\IEEEauthorrefmark{1}, Haoyuan Zhao\IEEEauthorrefmark{1}, Long Chen\IEEEauthorrefmark{1}, Xiaoqiang Ma\IEEEauthorrefmark{2}, Jiangchuan Liu\IEEEauthorrefmark{1}}

\IEEEauthorblockA{\IEEEauthorrefmark{1}School of Computing Science, Simon Fraser University, Canada}

\IEEEauthorblockA{\IEEEauthorrefmark{2}CSIS Department, Douglas College, Canada}

Emails: \{masamim, ycchou, hza127\}@sfu.ca; \{longchen.cs, mxqcs\}@ieee.org; jcliu@sfu.ca
}
\fi

\maketitle
\thispagestyle{plain}
\pagestyle{plain}

\begin{abstract}

Low Earth orbit Satellite Networks (LSNs) have been advocated as a key infrastructure for truly global coverage in the forthcoming 6G. This paper presents our initial measurement results and observations on the end-to-end network characteristics of Starlink, arguably the largest LSN constellation to date. Our findings confirm that LSNs are a promising solution towards ubiquitous Internet coverage over the Earth; yet, we also find that the users of Starlink experience much more dynamics in throughput and latency than terrestrial network users, and even frequent outages. Its user experiences are heavily affected by environmental factors such as terrain, solar storms, rain, clouds, and temperature, so is the power consumption. We further analyze Starlink's current bent-pipe relay strategy and its limits, particularly for cross-ocean routes. We have also explored its mobility and portability potentials, and extended our experiments from urban cities to wild remote areas that are facing distinct practical and cultural challenges. 

\end{abstract}

\blfootnote{A short version of this paper is to appear in IEEE INFOCOM 2023 (submitted on Aug. 1st, 2022; accepted on Dec. 1st, 2022)}

\section{Introduction}

\input{sections/intro}

\section{Related Work}
\label{sec:related_work}
\input{sections/related}

\section{System Setup and Methodology}
\label{sec:methods}
\input{sections/methods}

\input{sections/experiments}

\input{sections/experiments-wild}

\section{Further Discussions}
\label{sec:future_works}
\input{sections/future}

\section{Conclusion and Future Works}
\label{sec:conclusion}
\input{sections/conclusion}

\section*{Acknowledgement}
This project was supported by a Canada NSERC Discovery Grant and a British Columbia Salmon Recovery and Innovation Fund (No. 2019-045). The authors thank Dr. Will Atlas from the Pacific Salmon Foundation and Ian Clevenger from the Salmon Watersheds Lab at SFU, as well as the Heiltsuk First Nation, for their great support. We also thank the QQS (EYES) Projects Society for hosting us at the Koeye Lodge (in particular, the lodgekeepers Ian and Emily Files) and for their effort in protecting the natural environment of the Great Bear Rainforest and the Heiltsuk heritage.

\balance
\bibliographystyle{IEEEtran}
\bibliography{starlink_edge}

\end{document}

%% file: sections/intro.tex
Low Earth orbit (LEO) satellites operate at around 180 km to 2,000 km above the Earth surface, which, compared to the traditional Geosynchronous orbit (GEO) satellites at around 35,780 km, enable much shorter delays and much higher throughput for space-ground communications, albeit with smaller coverage.\footnote{\url{https://earthobservatory.nasa.gov/features/OrbitsCatalog}}
A large number of such LEO satellites, however, can form an {\em LEO Satellite Network} (LSN) constellation that collectively offers high-quality network services to ground users with truly global coverage. It has been suggested as a key infrastructure towards the upcoming 6G networking and beyond \cite{path_to_6g}. We have seen commercial deployment of LSN constellations in the past decade with rapidly growing attention from the general public. One of the industrial leaders \emph{OneWeb}, is building a constellation of 648 broadband satellites, which would eventually expand to 7,000.\footnote{\url{https://oneweb.net/resources/oneweb-streamlines-constellation}} Its rival, \emph{SpaceX's Starlink}, has launched more than 2,000 satellites to LEO and has received approval from the Federal Communications Commission (FCC) to bring that number up to 12,000.  FCC has further authorized Starlink to launch 7,500 generation 2 satellites in the near future.\footnote{\url{https://www.cnbc.com/2022/12/01/fcc-authorizes-spacex-gen2-starlink-up-to-7500-satellites.html}} The next-generation Starlink constellation may eventually harbor up to 30,000.\footnote{\url{https://www.space.com/spacex-starlink-satellites.html}}

\begin{figure}[t]
\centering
\includegraphics[width=0.8\linewidth]{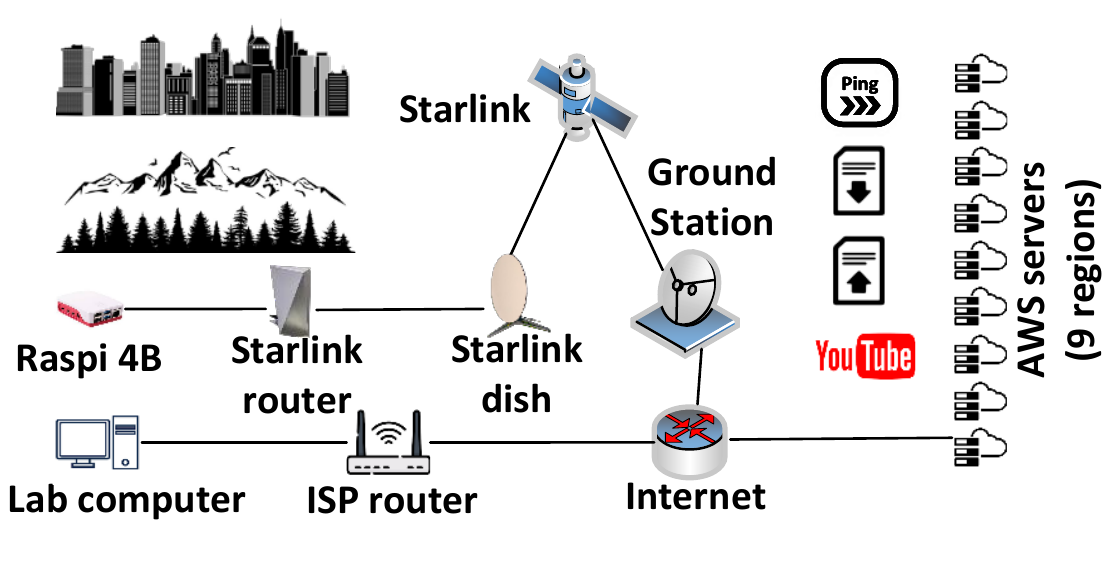}
\caption{Starlink measurement with diverse scenarios and tools.}
\label{fig:intro}
\end{figure}

Yet, LSNs remain in their early stages with limited user bases and user experiences. There are novel theoretical and simulation-based studies on satellite communications for both GEO and LEO. There exist, however, few comprehensive studies that have a deep insight into the practical performance of large-scale LSNs, except for scattered discussions in forums or with simulations \cite{kassing_exploring_2020, starperf_simulator}. The complex and dynamic topologies, as well as the heterogeneous and black-box architectures incurred during incremental deployment, further complicate how outstanding LSNs perform. To accelerate the design, deployment, and optimization of LSNs, a systematic measurement is indispensable, though  challenging.

In this paper, we present our initial results and observations on a systematic LSN measurement study. We focus on the Starlink network, which is arguably the largest LSN constellation to date, in terms of both the number of  satellites and the user base. We are particularly interested in the end-to-end network characteristics and performance with diverse configurations and applications. These are also the focus of most general users of the Internet, the mass consumer market that is targeted by Starlink. As a matter of fact, the Starlink network operator, SpaceX, offers  a plug-and-play black-box service to common end users, who can simply connect to a Starlink router through WiFi or Ethernet, without knowing the satellite communication details and the constellation operations. We are therefore interested in the following key questions:
\begin{itemize}
    \item Can today's Starlink achieve comparable performance as a typical modern terrestrial network does for end users?
    \item What are the key influential factors of the Starlink network services as perceived by end users?
    \item Has Starlink achieved global coverage through its constellation? If not, what are the challenges?
\end{itemize}

Our Starlink measurement, involving four dishes, has lasted over half a year starting from the beginning of 2022. We use scripts to automatically launch various tools and applications to communicate to servers distributed in different regions worldwide. We have performed field tests in urban cities and remote wild areas with a per-region collection of over 3 million total records. Our findings confirm that LSNs are a promising solution towards ubiquitous Internet coverage over the Earth; yet, we also identify a series of significant issues on the current Starlink's network services; in particular, the following observations:
\begin{enumerate}
    \item The throughput and latency experienced by a Starlink user are much more dynamic compared to terrestrial networks; there are frequent outages, too.
   
    \item  Starlink's user experience can be heavily affected by environments such as terrain, solar storm, rain, cloud, and temperature, so is its power consumption.

\end{enumerate}

We are also interested in the impact of the constellation's topology. Instead of establishing {\em Inter-Satellite Links} (ISLs) between satellites, Starlink is currently using a \emph{bent-pipe} strategy \cite{giuliari_icarus_2021, hauri_internet_2020}, where a LEO satellite must relay the traffic to a \emph{Ground Station} (GS) for further routing \cite{fischer_predictable_2013}, as shown in Fig. \ref{fig:intro}. We have validated that the bent-pipe is generally of one hop and involves the closest GS only  before switching to the terrestrial network, instead of multiple. We analyze this strategy and its potential limits, particularly for cross-ocean routes.

We also examine Starlink's mobility and portability support. Even though throughput does not change much during movement, our experiments suggest that mobility with Starlink still requires significantly more work as outages happen frequently with high latency. We further extend our experiments from urban cities to the wild remote areas which reveal a series of issues regarding  global coverage, including unique practical and cultural challenges.

The remainder of this paper is organized as follows. The related works are reviewed in Section \ref{sec:related_work}, followed by the measurement methodology in Section \ref{sec:methods}. We then present and analyze the measurement results in an urban city in Section \ref{sec:starlink_in_major_city} and those in the wild in Section \ref{sec:starlink_in_wild}, respectively.  We discuss a series of advanced issues in Section \ref{sec:future_works} and conclude the paper with future directions in Section \ref{sec:conclusion}.

%% file: sections/related.tex
\begin{figure}[t]
    \centering
    \subfigure[Starlink's first shell generated by Hypatia \cite{kassing_exploring_2020}.]{
        \includegraphics[width=.3\linewidth]{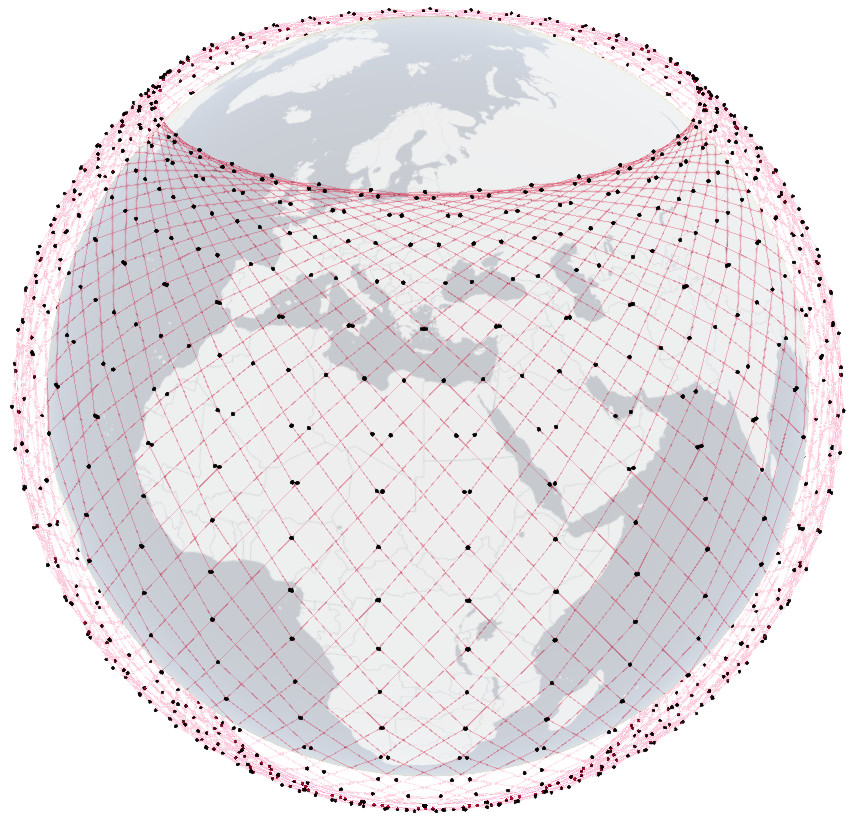}
        \label{fig:starlink_structure_hypatia}
    }
    \subfigure[Starlink's network structure.]{
        \includegraphics[width=.5\linewidth]{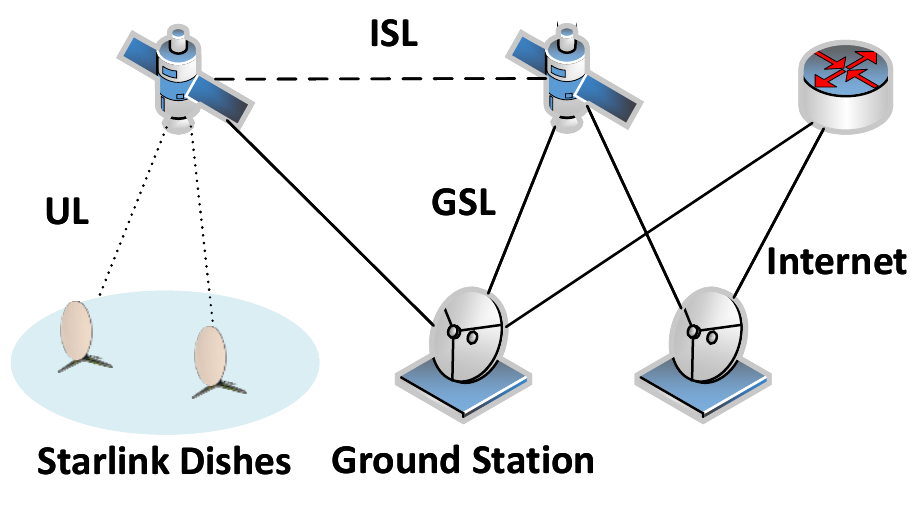}
        \label{fig:starlink_structure_network}
    }
    \caption{(a) The constellation of the Starlink's first shell; (b) The Starlink's network structure with ISLs, GSLs, and ULs.}
    \label{fig:starlink_structure}
\end{figure}

There have been many pioneering works on satellite communications, from theoretical analysis \cite{handley_delay_2018, bhattacherjee_gearing_2018, giuliari_internet_2020, time_graph_chen_2021}, to system design \cite{giuliari_icarus_2021, hauri_internet_2020, denby_orbital_2020, bhattacherjee_network_2019, lai_spacertc_2022}, and to practical deployment \cite{handley_using_2019, vasisht_l2d2_2021, singh_community-driven_2021}. Early measurement on real-world practical systems or simulations have focused on GEO satellite networking \cite{henderson_protocol_sat_1999, caini_tcp_geo_2009}. Ground-to-satellite distance and hence the signal latency is a physical barrier that can only be alleviated by moving to the LEO. In the past decade, LEO constellations have received much attention as a means towards ubiquitous Internet connectivity beyond 5G. There have been extensive studies on the interaction between LEO satellites to ground stations \cite{vasisht_l2d2_2021, singh_community-driven_2021, orbit_cast_lai_2021, wang_enhancing_2022, zhang_enabling_2022}, constellation topology management, addressing and routing \cite{giuliari_internet_2020} \cite{cyber_physical_li_2021}, as well as the potential impact of environmental factors, such as rain \cite{vasisht_l2d2_2021} and solar storms \cite{jyothi_solar_2021}. 

Recent years have seen real-world deployments like SpaceX's Starlink, OneWeb, and Telesat. There have been numerous articles and posts on these LSNs, in particular, Starlink, from newsmedia and user forums,\footnote{\url{www.reddit.com/r/Starlink/comments/jllpet/starlink_beta_report_specifications_dimensions/}} most of which were simple reviews. Studies on LEO constellation networks have largely relied on simulators, such as \texttt{Hypatia} \cite{kassing_exploring_2020} and \texttt{StarPerf} \cite{starperf_simulator}. They are good for model and design validation, but real-world measurement for such large-scale complex systems as Starlink is irreplaceable. We have seen earlier measurements \cite{reed_darksat_2020, reed_darksat_2021, uran_starlink_2021, khalife_tracking_2022, neinavaie_starlink_tracking_2022}, which were conducted with relatively simple setups and limited dynamics of influential factors, and often focused on physical layer behaviors. There have been recent measurement works with Ookla's speedtest\footnote{\url{https://www.speedtest.net/}} and a browser extension\cite{michel_first_2022, kassem_browser-side_2022}. They have examined QUIC performance and browser latency with Starlink. Our work, in parallel time as theirs, attempts to provide a systematic measurement on Starlink's performance from  the  common end users' perspective. Our measurement has a more global coverage with distant endpoints, and considers video streaming, remote wild northern areas, and power consumption.

%% file: sections/methods.tex
\subsection{Starlink Network Overview}

SpaceX's Starlink is a prominent representative of the state-of-the-art LEO satellite networks, serving over 400,000 subscribers worldwide as of today. Its system consists of three main components: a constellation of LEO satellites, a network of ground stations, and user terminals. The   constellation currently has over 2,000 satellites at different LEO groups, known as {\em shells} \cite{kassing_exploring_2020, hauri_internet_2020}. Most of the deployed satellites are in the first shell, which contains  72 orbits at an altitude of 550 km in planes inclined 53\textdegree{}, each accommodating a maximum of 22 satellites (see Fig. \ref{fig:starlink_structure_hypatia}).

\begin{figure}[t]
    \centering
    \def\svgwidth{1\linewidth}
    \formatfig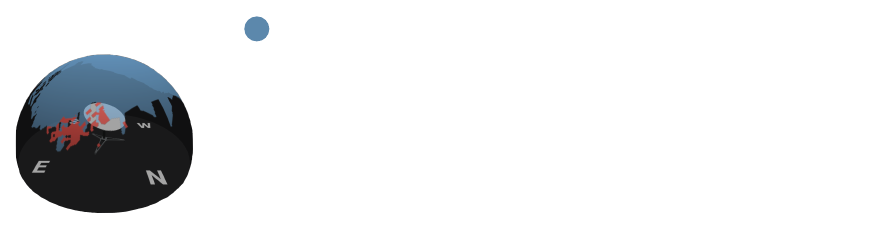\formatnormal
    \caption{Samples of Starlink visibility maps, together with obstruction ratios taken from the Starlink app logs. Dishes A and B are in Burnaby; dish F is in \blind{a northern remote area}\normal{Koeye Point}.}
    \label{fig:sami_obstr_map}
\end{figure}

The satellites employ phased array beamforming and digital processing in the Ku- and Ka-bands for data communications \cite{fcc_attachment_2016}. At the ground, instead of directly connecting from handsets to satellites, a dish (nicknamed as {\em Dishy McFlatface}) is needed to communicate through a {\em User Link} (UL) with a LEO satellite that is currently visible and accessible, and the dish then serves local user devices. Two generations of dishes are available for consumers: the round-shaped Gen 1 and the rectangular Gen 2. Gen 1 spans 23.2" in diameter and weighs in at 7.3 kg, whereas Gen 2 is slightly smaller (19" x 12") and lighter (4.2 kg) with more bandwidth streams in the router (3x3 MU-MIMO versus 2x2 in Gen 1). The latter has better portability but remains bulky. Not only the sky needs to be clear, but the setup location needs to be wide enough to accommodate the dish size and any movement required for its self-adjusted tilt. A visibility map provided by the Starlink app can help with the placement (see Fig.~\ref{fig:sami_obstr_map} for an example), which also gives a numerical {\em obstruction ratio} in its debug mode. We have used both dishes in our measurement and found that their network services experienced by end users are largely similar, though Gen 2 seems more friendly to TCP. 

Different from early satellite services that often serve a niche market, Starlink targets the next generation broadband Internet service for the global consumer market.  To this end, it provides end users a plug-and-play Internet access interface that hides the aforementioned technical details of satellite communications. The user kit consists of three main components: the dish, a power-over-Ethernet power supply adapter, and a router with WiFi and Ethernet port (Gen 1) or WiFi only (Gen 2) for local devices. The particular satellite connecting to the dish at a time will act as a repeater, relaying the signal through the Ka-band to a {\em Ground Station} (GS) through a {\em GS-to-Satellite Link} (GSL), which further connects to the global Internet, or vice versa (see Fig.~\ref{fig:starlink_structure_network}). Currently, there are at least 65 GSes located in North America and 23 GSes in Europe, as well as others scattered worldwide.\footnote{\url{https://starlink.sx/}}   To successfully establish this bent-pipe for traffic relay, the GS and the dish  must be both in the field of view of the satellite at the same time. Therefore, their distance should not exceed about 1,000 km. Also note that a LEO satellite has a fly-over dwell time of only around 10 minutes over one spot on the ground \cite{vasisht_l2d2_2021}; when it moves out, the service needs to be handover to a new satellite moving into the area (if there is). Starlink has started laser-based ISL (Inter-Satellite Link) experiments; yet large scale implementation will not happen till the end of 2022 \cite{garreffa_spacex_2022}.

\subsection{Measurement Topology and Environment}

Our measurement study started in January 2022 and lasted 7 months. We have deployed 4 dish kits in different locations for data generation, communication, and collection. Among them, three are Gen 1 (referred to as A, B, and C, respectively) and one is Gen 2 (referred to as F) which is easier to carry to remote areas. Our experiments span over a wide range of terrains, including major cities, namely Vancouver (Burnaby and Coquitlam in the metro-Vancouver, in particular) in British Columbia (BC), Canada, with open sky spaces, remote areas with steep valleys or at the far north (Koeye Point, BC, which is near the current service boundary of Starlink), temperate rain forests with heavy obstacles, etc. Dish A and B are stationed in Burnaby whereas dish C is setup in Coqutilam, and Dish F is stationed at Koeye Point. To collect the measurement data with minimum signal, workload interference, and power demands, a Raspberry Pi 4B (Raspi) was directly connected to each Starlink router through an Ethernet cable, except for dish F which  has a WiFi interface only.  We have performed measurements to 9 destination servers deployed using Amazon Web Services (AWS), which are geo-distributed worldwide: Sao Paulo, Singapore, Sydney, North California, Bahrain, Tokyo, London, Cape Town, and Mumbai. This overall setup can be seen in Fig.~\ref{fig:intro}. 

For the experiments where domestic terrestrial Internet access is available, we have used a PC with an i7-10700KF CPU and the same AWS servers as the baseline for comparison. The baseline presented in this paper is running over a terrestrial Cable service with a maximum download speed of 800 Mb/s. We have experimented with different terrestrial broadband Internet services of similar maximum speeds, and have observed similar baseline performance.  We also used an Emporia smart plug\footnote{\url{https://www.emporiaenergy.com/emporia-smart-plug}} to monitor the instant and historical power consumption of the dish kit. To understand the impact of weather conditions, we extract hourly data from a climate station at Vancouver Harbour, which is approximately 10 km away from dish A.

\subsection{Target Applications and Measurement Tools}

Our measurement cover different protocols and applications, from standard Web browsing and  file transfer to high-demand video streaming. We use \texttt{iperf3} to measure the TCP and UDP throughput in the transport layer. We used cubic TCP as the default, though further studies would be conducted to compare with other variants.  For each destination server,  two \texttt{iperf3} are launched at different ports for  downloading and uploading, respectively, allowing simultaneous measurements of both. We also use the {\em Secure Copy Protocol} (SCP) to copy files with different data sizes bi-directionally to the AWS regions.  \texttt{fallocate} has been used to create files with data sizes ranging from 1 to 300 MB.  \texttt{ping} and \texttt{traceroute} have been used to track and analyze the network path characteristics and routing strategies. We have used the Starlink Mobile App and YouTube's built-in tool \texttt{Stats for nerds} to collect related network, user, and application information. We have designed Python-based scripts to automatically launch the tests and collect the output data, which will be made publicly available.

%% file: sami_starlink_obstructions.pdf_tex
\begingroup%
  \makeatletter%
  \providecommand\color[2][]{%
    \errmessage{(Inkscape) Color is used for the text in Inkscape, but the package 'color.sty' is not loaded}%
    \renewcommand\color[2][]{}%
  }%
  \providecommand\transparent[1]{%
    \errmessage{(Inkscape) Transparency is used (non-zero) for the text in Inkscape, but the package 'transparent.sty' is not loaded}%
    \renewcommand\transparent[1]{}%
  }%
  \providecommand\rotatebox[2]{#2}%
  \newcommand*\fsize{\dimexpr\f@size pt\relax}%
  \newcommand*\lineheight[1]{\fontsize{\fsize}{#1\fsize}\selectfont}%
  \ifx\svgwidth\undefined%
    \setlength{\unitlength}{251.28000069bp}%
    \ifx\svgscale\undefined%
      \relax%
    \else%
      \setlength{\unitlength}{\unitlength * \real{\svgscale}}%
    \fi%
  \else%
    \setlength{\unitlength}{\svgwidth}%
  \fi%
  \global\let\svgwidth\undefined%
  \global\let\svgscale\undefined%
  \makeatother%
  \begin{picture}(1,0.28653295)%
    \lineheight{1}%
    \setlength\tabcolsep{0pt}%
    \put(0,0){\includegraphics[width=\unitlength,page=1]{sami_starlink_obstructions.pdf}}%
    \put(0.32421177,0.24306103){\makebox(0,0)[lt]{\lineheight{1.25}\smash{\begin{tabular}[t]{l}Clear View\end{tabular}}}}%
    \put(0.57141465,0.24306103){\makebox(0,0)[lt]{\lineheight{1.25}\smash{\begin{tabular}[t]{l}Obstructions\end{tabular}}}}%
    \put(0,0){\includegraphics[width=\unitlength,page=2]{sami_starlink_obstructions.pdf}}%
    \put(0.11931477,0.00733051){\makebox(0,0)[t]{\lineheight{1.25}\smash{\begin{tabular}[t]{c}Dish A: 2.7\%\end{tabular}}}}%
    \put(0,0){\includegraphics[width=\unitlength,page=3]{sami_starlink_obstructions.pdf}}%
    \put(0.3766171,0.0058469){\makebox(0,0)[t]{\lineheight{1.25}\smash{\begin{tabular}[t]{c}Dish B: 4.7\%\end{tabular}}}}%
    \put(0.63391939,0.0058469){\makebox(0,0)[t]{\lineheight{1.25}\smash{\begin{tabular}[t]{c}Dish C: 24.9\%\end{tabular}}}}%
    \put(0,0){\includegraphics[width=\unitlength,page=4]{sami_starlink_obstructions.pdf}}%
    \put(0.89122168,0.00734217){\makebox(0,0)[t]{\lineheight{1.25}\smash{\begin{tabular}[t]{c}Dish F: 0.0\%\end{tabular}}}}%
  \end{picture}%
\endgroup%

%% file: sections/experiments.tex
\input{tables/throughput_mean_gaps}

\begin{figure}[t]
    \centering
    \includegraphics[width=0.8\linewidth]{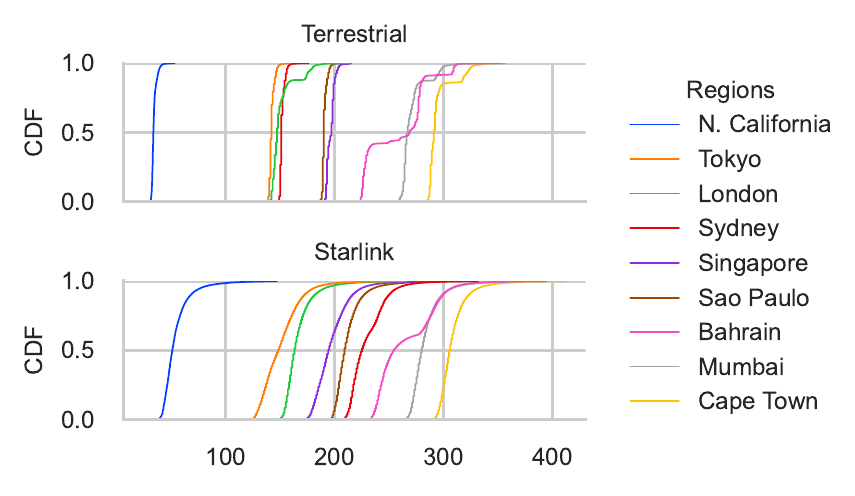}
    \caption{Cumulative distribution of latency (ms) globally.}
    \label{fig:no_outliers_ping_cdf}
\end{figure}

\begin{figure}[t]
    \centering
    \includegraphics[width=0.8\linewidth]{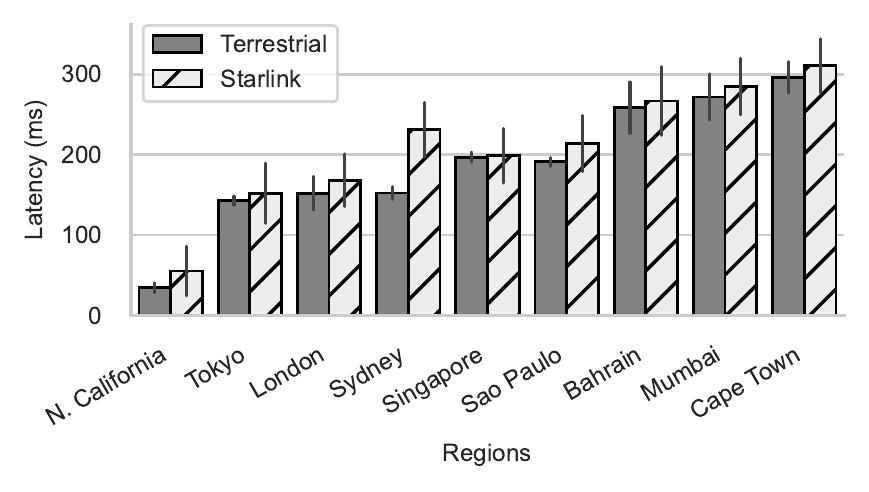}
    \caption{Average latency in different regions. Vertical lines correspond to standard deviations of data samples.}
    \label{fig:ping_avg_hist}
\end{figure}

\section{Urban City Measurement: Starlink versus Terrestrial Networking}
\label{sec:starlink_in_major_city}

We start our measurement in Vancouver, a major city at the Pacific West Coast with all the state-of-the-art terrestrial Internet infrastructures and Starlink coverage. We have deployed the dishes in different locations in the city's residential areas with sufficient open surrounding spaces, and hence should reveal the best performance of Starlink.

\subsection{End-to-End Latency}

We have measured the latency across 9 AWS regions worldwide and looked at the difference between Starlink and the terrestrial network's latency. In summary, we have found that Starlink's latency is slightly higher (10\%) than that of the terrestrial network on average and is much more unstable (see Fig.~\ref{fig:no_outliers_ping_cdf},~\ref{fig:ping_avg_hist}). Obstruction, satellite movements, and ISP routing decisions are potential factors that affect Starlink's latency.

\begin{figure}[t]
    \centering
    \includegraphics[width=0.8\linewidth]{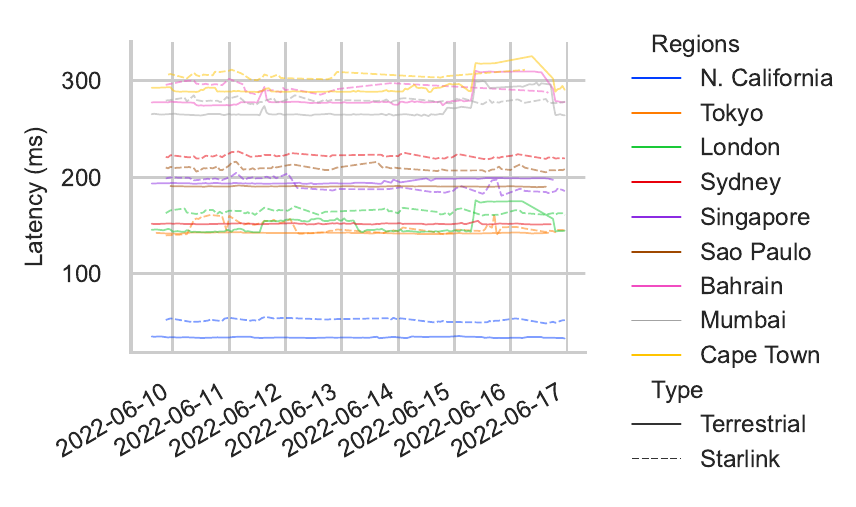}
    \caption{Hourly latency over time during a one week period.}
    \label{fig:hour_ping}
\end{figure}

\begin{figure}[t]
    \centering
    \def\svgwidth{0.8\linewidth}
    \includegraphics[width=0.8\linewidth]{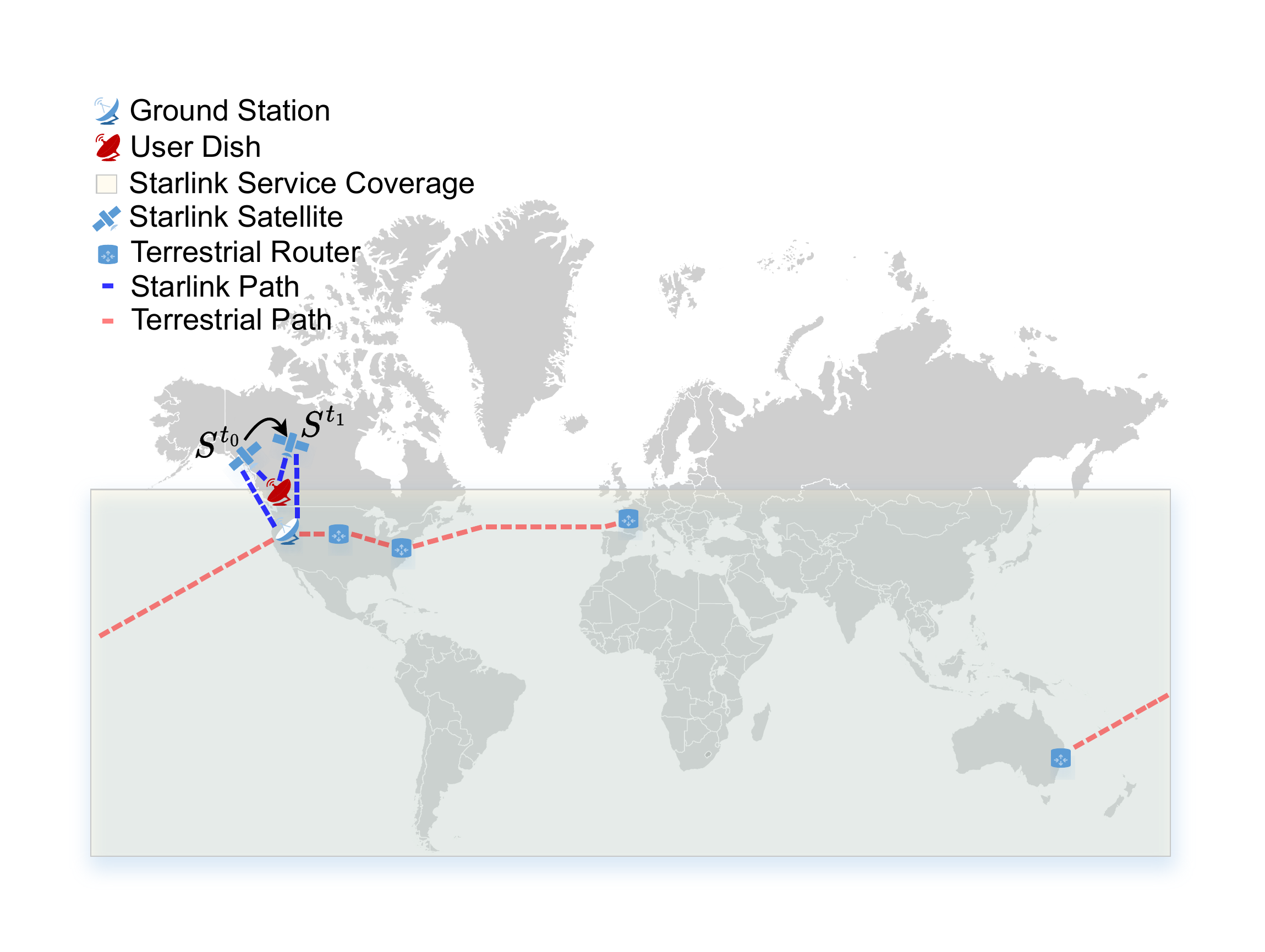}
    \caption{Current Starlink network routing framework.}
    \label{fig:starlink_current_routing}
\end{figure}

\subsubsection{Latency across Regions}

We can see noticeable gaps between the latencies of Starlink and that of the terrestrial network in Fig.~\ref{fig:no_outliers_ping_cdf} with the average gaps shown in TABLE~\ref{table:lat_perc}. The gaps typically range from 1.8 to 22.8 ms away from the terrestrial network, with Sydney being an outlier (3.4x to 43.6x larger gap). There is a noticeable increase in latency for Bahrain, but since the terrestrial network also observes such an increase, the dish is likely not the culprit. This could be due to a connection issue with the AWS server at Bahrain. Starlink's latency being almost always just slightly above that of the terrestrial network suggests that only a small number of hops increase this latency, coinciding with the idea of the single bent-pipe transmission framework, i.e., only the first hop uses a bent-pipe \cite{handley_using_2019}. Otherwise, if multiple bent-pipe transmissions are involved, there would be a larger variation of the latency discrepancy between Starlink and the terrestrial network, and Starlink would have lower latency to some destinations. Sydney being an outlier also reaffirms this. When Starlink started accepting requests for a Maritime version of their dish\footnote{\url{https://www.starlink.com/maritime}} in July 2022, its coverage map only spans across the coastal waters around in-land regions. A switch to a terrestrial network would be required to cross the Atlantic or Pacific oceans through  submarine fibre.  For example, to communicate with Sydney, Starlink currently chooses to route packets through a \normal{Seattle}\blind{US West} AWS server after the first hop bent-pipe. 

\subsubsection{Latency Variation}
Starlink's latency has a very high variation, around 3.8 times that of the terrestrial network (Shown in Fig.~\ref{fig:ping_avg_hist}). Given that the terrestrial network keeps a relatively stable latency as seen in Fig.~\ref{fig:hour_ping}, the common rise and falls for Starlink not observed in the terrestrial network are likely due to the constant network path changes as satellites move closer or further away from the dish, affecting the latency of the single bent-pipe transmission.  As shown in Fig.~\ref{fig:starlink_current_routing}, when a connected satellite S is moving away from the user dish (from $S^{t_0}$ to position $S^{t_1}$, where the superscripts $t_0$ and $t_1$ represent the satellite S at different times), the latency would increase continuously until a handover to a new satellite happens. This is also validated by a closer look at the arrival time at the packet level --- the jitters of Starlink are 0.255 ms (download) and 2.715 ms (upload), both of which are over 5x higher than that of the terrestrial network (0.041 ms for download and 0.546 ms for upload, respectively). 

\begin{figure}[t]
    \centering
    \subfigure[Download throughput]{
        \includegraphics[width=0.8\linewidth]{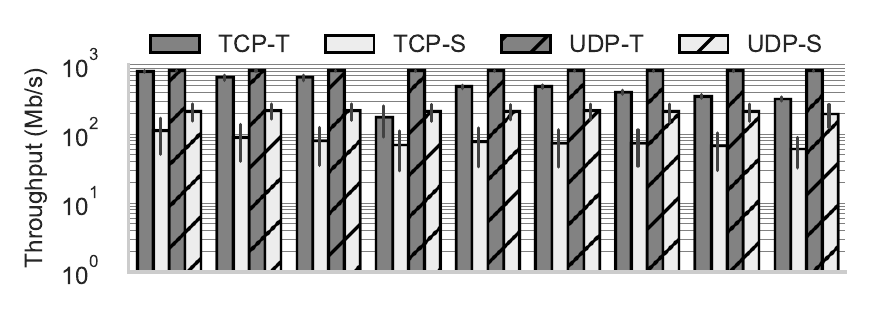}
    }
    \\
    \subfigure[Upload throughput]{
        \includegraphics[width=0.8\linewidth]{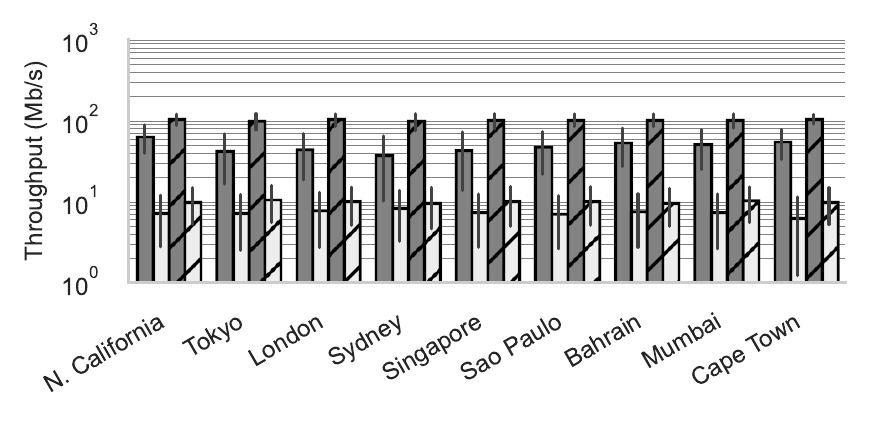}
    }
    \caption{Throughput measurements across different regions for the terrestrial (TCP-T/UDP-T) and Starlink (TCP-S/UDP-S) network, respectively. Vertical lines correspond to standard deviations of data samples.}
    \label{fig:throughput_summary}
\end{figure}

\subsection{End-to-End Throughput}

We have measured the throughput of the Starlink dish using \texttt{iperf3} with both TCP and UDP. As shown in Fig.~\ref{fig:throughput_summary}, Starlink achieves good throughput ($\sim$80 Mb/s on average) but again lacks stability. The  standard deviations of throughput for Starlink and terrestrial networks are 50.71\% and 34.44\%, respectively. Again, this variation is likely due to constantly changing network paths from satellite movements and handovers, in addition to potentially low-cost terrestrial network routing after being passed over from the Starlink network.

For UDP measurements, we gradually increase \texttt{iperf3}'s traffic flow until the path is saturated. Starlink's achieved throughput is similar across all regions (see  TABLE~\ref{table:reg_avg_throughput}). This suggests either the dish's hop is physically the bottleneck or traffic engineering has been applied to this hop at UL or GSL. The latter would be the root cause as the results in \ref{subsubsec:Precipitation_Temperature_Dish_Power} suggest that the throughput may be throttled for power consumption control. We have also observed bursty packet losses that are much higher than the terrestrial network. Such losses are not necessarily caused by congestion, as we have observed them with a low UDP throughput experiment as well. For 20\% of the presented throughput, which is low enough to mitigate congestion factors of packet loss \cite{xu_understanding_2020}, we still see bursty losses of 0.24\% and 1.24\% on average for downloads and uploads, respectively. This is consistent with early simulation studies for satellite communications \cite{shao_noma-based_2019}. Interestingly, there is also a cycle around 12 hours. The average and peak values differ from cycle to cycle but are similar across regions during each cycle, suggesting that the losses happen most likely on the bent-pipe, in particular, the UL, which will be further examined in our analysis of the routing strategy in Section~\ref{sec:routing_strat}.
 
For TCP traffic, Starlink has experienced more issues with congestion control than the terrestrial network. Our results show that Starlink has a {\em bandwidth utilization} \cite{xu_understanding_2020} of 39.0\% , which is noticeably lower than that of the terrestrial network (46.8\%), suggesting TCP's congestion control is quite sensitive to the dynamics in the satellite hop. The newer Gen 2 dish may optimize TCP uploads with a 1.76x more throughput as compared to the Gen 1 dishes in the urban city. Unfortunately, UDP uploads do not see a similar increase, which is 0.73x of the Gen 1 dishes in terms of throughput. This suggests that the improvement of TCP uploads could be due to a more optimized access method in the updated dish or the revised router design.

\input{tables/2022-06-05_end_avg_throughput}

\input{sections/experiments/scp}

\begin{figure*}[t]
    \centering
    \includegraphics[width=0.8\linewidth]{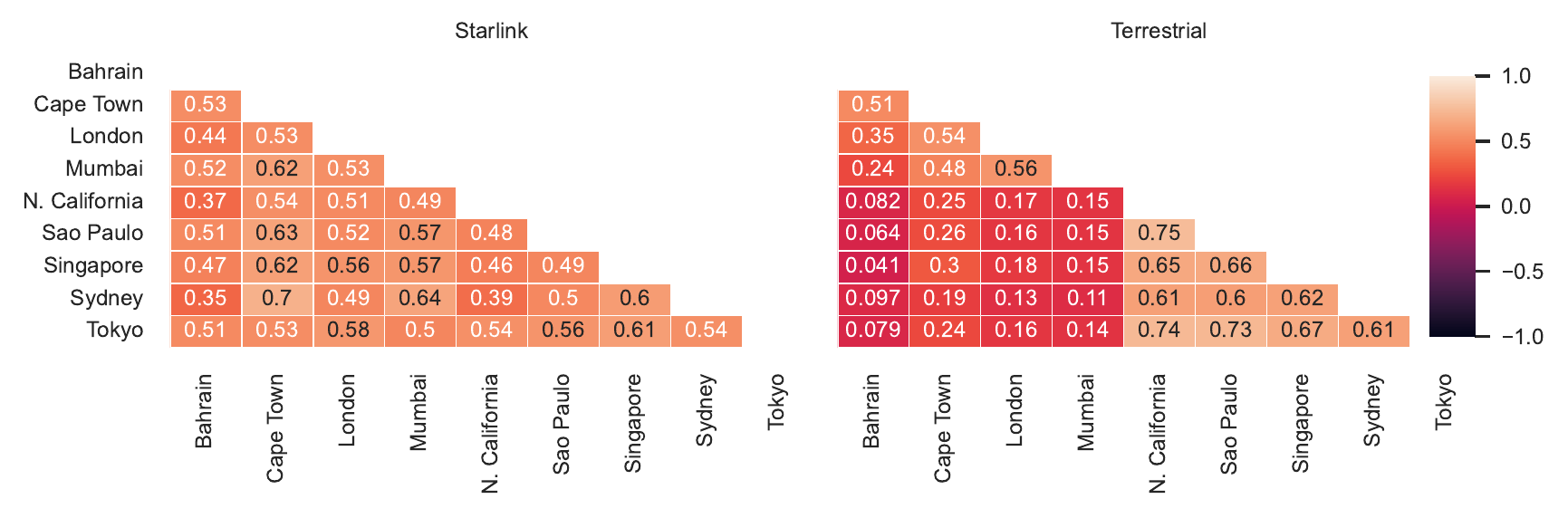}
    \caption{Pearson latency correlation heatmap between regions.}
    \label{fig:latency_reg_corr}
\end{figure*}

\subsection{Routing Strategy}
\label{sec:routing_strat}

The results from our \texttt{traceroute} measurements suggest that the Starlink network only does one bent-pipe communication (e.g., the first UL and GSL) along the route and then enters the terrestrial network to arrive at the destination, or vice versa. In our experiments, \emph{the Starlink network will always connect to the geographically closest GS from the dish no matter where the destination is}. We only see a single SpaceX Services ISP entry located at a GS nearest to the dish before the routing switches to a terrestrial ISP. To the best of our efforts, we have found no reference of Starlink using L2 routing so far. As shown in Fig.~\ref{fig:hour_ping}, the regions, especially Sydney, Sao Paulo, and London have similar latency patterns for the Starlink network. We use the Pearson correlation  between all the regions to further analyze this point.\footnote{Pearson correlation works by drawing a best fit line through the datapoints on a plot and providing a coefficient that describe the distribution from the line to denote the correlation between two variables. The closer the Pearson coefficient is to 1, the higher the correlation between the two variables. See \url{https://statistics.laerd.com/statistical-guides/pearson-correlation-coefficient-statistical-guide.php}}  The correlation matrices in Fig.~\ref{fig:latency_reg_corr} show that all the regions the dish communicates with have moderate correlations, further suggesting they have similar latency patterns. Despite communicating with different regions, both observations suggest that similar satellites, GSes, and routes have been  used. In other words, this {\em single-bent-pipe architecture} does not participate in the routing strategy other than serving as the first hop to bridge to the Internet (or last hop should the Starlink user be the destination). Otherwise, considering that London is closer to US East, the traffic from our measurement location (at the Pacific
West Coast)  should be routed by Starlink to GSes in US East using multiple bent-pipe transmissions before switching to the terrestrial network, which should have lowered the correlations against other regions not in the same general direction as London, such as N. California, Sydney, and Tokyo. Unfortunately, this is currently not the case as the correlations shown in Fig. \ref{fig:latency_reg_corr} are moderate.

So far our discussion has been limited to the case of only one end being a Starlink user. When Starlink becomes a ubiquitous Internet service, frequently, both network endpoints would be dish users. Ideally, one would expect that, when two nearby dishes are both within the service area of a satellite, they can {\em peer-to-peer} exchange data using the satellite to relay their respective ULs, without going through GSLs or other terrestrial Internet nodes. To verify this, we have placed dishes A and B together, attempting to connect them directly. Unfortunately, we have discovered that only \texttt{ssh} connections are allowed by Starlink in this setup. In fact, Starlink prevents users from editing more advanced features in typical routers such as port forwarding. In the end, an external proxy must be used, e.g. through \texttt{remote.it},\footnote{\url{https://www.remote.it/resources/how-to-port-forward-on-starlink}} which, being customized for Starlink, sets up an external AWS server as the public proxy. The throughput averaged 2-3 Mb/s with occasional bursts to 9-12 Mb/s; however, 27\% of the throughput readings are simply zero. Clearly, the tunnel-based solution does not work well yet, needing improvements for pair-wise dish communications.

\subsection{Environmental Influential Factors}

\subsubsection{Obstruction}

The Starlink App gives the sky visibility map of the dish at the current location (Fig.~\ref{fig:sami_obstr_map}), together with the obstruction status. The latter is only visible in the debug mode of the App through \texttt{ObstructionStats} $\rightarrow$ \texttt{CurrentlyObstructed} and \texttt{FractionObstructed}. We referred to the fraction as {\em obstruction ratio} in this paper, which is not always correlated with the visibility map. It takes around 12-24 hours to build the initial map, but the obstruction ratio is updated more frequently to capture live dynamics, e.g., fallen leaves that cover the dish. We have found that, even if a location has a good visibility map, a transient obstruction due to foreign objects can lead to spikes in latency or even network outage, implying that the dish is very sensitive to surface obstructions.\footnote{Though the Starlink App conceals obstruction status in its debug mode with no further details, certain information can be extracted from the debug log through a third-party tool {\texttt Obstruction Viewer} (\url{http://starlink.dsmfaq.com/tools/obstructions.html}).} In short, Starlink's end user needs both a clear sky and a clean dish to achieve the best performance.

\subsubsection{Solar and Geomagnetic Storms}

Starlink's throughput can be heavily affected by solar and geomagnetic storms. A solar storm, or Coronal Mass Ejection (CME), is when massive magnetized particles from the Sun are ejected in a specific direction, e.g., towards the Earth, which can interact with the earth's magnetic field \cite{jyothi_solar_2021}.\footnote{\url{https://www.swpc.noaa.gov/phenomena/geomagnetic-storms}} We have found that, during February 3rd and 4th, 2022, the dish's throughput dropped dramatically from 100 Mb/s to 5 Mb/s. This coincides with two events: A SpaceX mission that launched 49 more satellites to space on Feb. 3rd and a geo/electro-magnetic storm on Feb. 4th that forced 40 satellites to re-enter the Earth's atmosphere.\footnote{\url{https://www.spacex.com/updates/\#sl-geostorm}} The storm may have strongly interfered with the UL/GSL similar to how GPS radio waves were affected by the change in atmosphere due to a previous geomagnetic storm \cite{linty_effects_2018}. Another reason could be that the satellite constellation may have been put in a less serviceable maintenance mode, leading to poor communication performance. Interestingly, uploads do not seem to be affected as the throughput stayed relatively stable. This implies that the two directions are not symmetric when facing such storms, which can be an important factor to consider when integrating Starlink and the terrestrial network infrastructures.

\begin{figure}[t]
\centering
\includegraphics[width=0.8\linewidth]{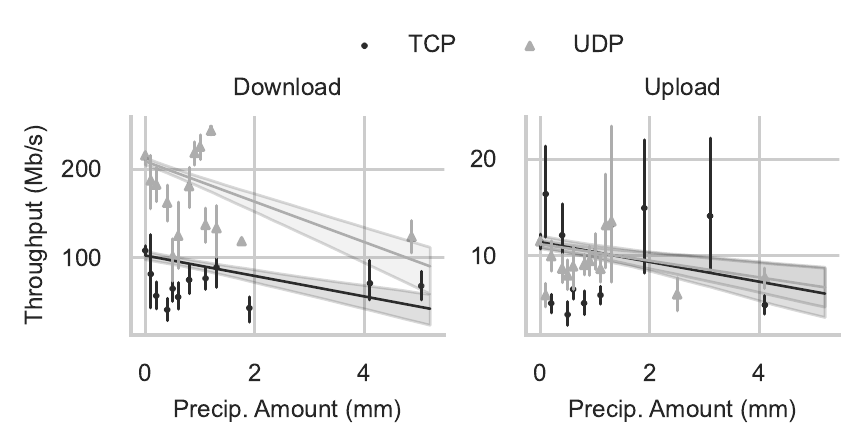}
\caption{Starlink throughput with precipitation (vertical lines correspond to 95\% confidence interval; translucent area outlines regression estimates).}
\label{fig:thr_precip}
\end{figure}

\subsubsection{Precipitation, Temperature, and Dish Power}
\label{subsubsec:Precipitation_Temperature_Dish_Power}

We have observed that the Starlink's throughput can be significantly affected by the weather. The results imply inverse correlations for throughput in comparison to precipitation and temperature.

Throughput drops on average 27\% during any precipitation (see details in Fig.~\ref{fig:thr_precip}). We have also found that the throughput is capped with heavy rain ($>4$ mm per hour).\footnote{\url{https://water.usgs.gov/edu/activity-howmuchrain-metric.html}} Downloads are affected more than uploads with UDP download having the maximum inverse correlation of 0.34. For instance, UDP downloads is on average 215 Mb/s with no precipitation and almost halves to around 120 Mb/s with 4.1 to 5.2 mm of precipitation. Rain attenuation is the likely culprit for the drop in throughput as Ka- and Ku-band radio waves can suffer heavily in rainfall \cite{panagopoulos_satellite_2004, kourogiorgas_space-time_2018}. Additionally, clouds that form in the sky could also interrupt data communications with satellites. It is known that even a light cloud can affect the satellite signal strength by around 10\% \cite{vasisht_l2d2_2021}. Thick clouds are usually coupled with heavy rainfall, further blocking the network paths in ULs and GSLs.

We have also observed a 5-26\% drop in throughput when the temperatures are above 12$ ^\circ $C. The Starlink dish does not have an active cooling mechanism, only passively cooling through its aluminum back-plate,\footnote{\url{https://www.vice.com/en/article/7kvkga/spacexs-satellite-internet-shuts-down-because-its-too-hot-in-arizona}} which can be a significant challenge when used in the summer or tropics.
 
Using an Emporia smart plug, we find that a current dish's power consumption averages around 56.3 Watts but can go as high as 144.5 Watts. This is considerably lower than an earlier reported measurement of 105 Watts on average and 190 Watts maximum \cite{uran_starlink_2021}. Such power saving is likely done through firmware optimization as the dishes' form factors remain the same, and it is certainly helpful for heat control. As an extra experiment during a snow storm in Burnaby at night from Dec. 19 to 20, 2022, Starlink enters a {\em snow melting} mode for 10 hours, increasing its power consumption to an average of 146.3 Watts with a maximum of 188.6 Watts. If Starlink is used in cold and snowy climates, the user must be careful as the dish requires significantly more energy to melt the snow that could accumulate. This feature however can be turned off in the Starlink app if  the user has alternative methods of keeping the Starlink clear from snow.

Since the dish kit is a blackbox, we cannot directly measure the power consumption of individual modules. However, we compared the power consumption against throughput and find that there is no obvious correlation between them, which contrasts with  the intuition that higher throughput requires more power. In fact, in some cases, we have seen that they are inversely correlated for both upload and download. These suggest that the dish does not necessarily amplify the signal further for higher throughput and that the dish may throttle the maximum achievable throughput to control the power consumption. Consider the summary in TABLE~\ref{table:reg_avg_throughput}, the highest average TCP throughput in our experiments is around 108 Mb/s for download and 6 Mb/s for upload; yet a previous experiment in Austria has observed much higher throughput for both directions: 175 Mb/s for download and 17 Mb/s for upload  \cite{uran_starlink_2021}. This could be due to obstructions and coverage differences; however, the Starlink software update to lower power consumption may have also affected the throughput.\footnote{\url{https://www.tesmanian.com/blogs/tesmanian-blog/ukraine-starlink}} An updated future measurement in the same environment would be required to test this conjecture.

\begin{figure}[t]
    \centering
    \includegraphics[width=0.8\linewidth]{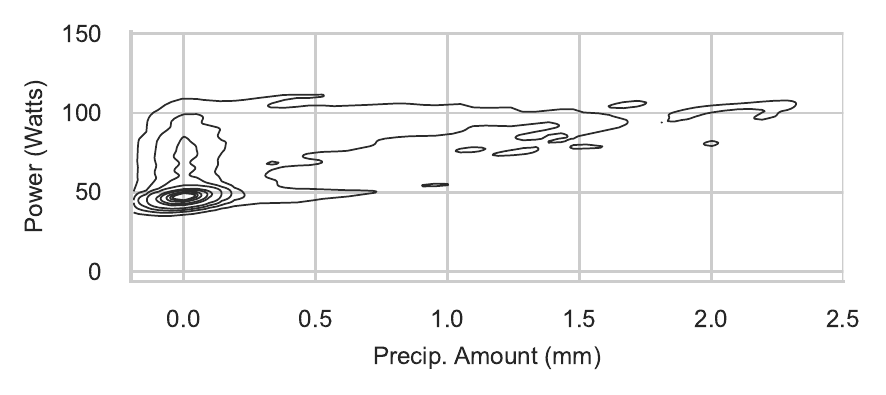}
    \caption{Power consumption versus hourly precipitation.}
    \label{fig:watt_precip_corr}
\end{figure}

On the other hand, as shown in Fig.~\ref{fig:watt_precip_corr}, the dish's power consumption is correlated with precipitation. Without rain, power consumption is lower in general, with occasional spikes due to other factors such as tuning  the dish direction or establishing  a UL with farther away satellites. When there is rain or heavy clouds, power consumption becomes persistently higher, likely in an attempt to deal with the interference from the rain or clouds.

%% file: tables/throughput_mean_gaps.tex
\begin{table*}[t]
\begin{center}
\caption{Average Latency Gaps between Starlink and Terrestrial Networks}
\label{table:lat_perc}
\begin{tabular}{|c|rrrrrrrrr|}
\hline
& Bahrain & Cape Town & London & Mumbai & N. California & Sao Paulo & Singapore & Sydney & Tokyo \\
\hline
 & \multicolumn{9}{c|}{$(\mu_s - \mu_t)^{\mathrm{*}}$} \\
\hline
Gap & 8.4 & 14.3 & 16.0 & 12.9 & 20.9 & 22.8 & 1.8 & 78.4 & 9.2 \\
\hline
\multicolumn{10}{l}{$^{\mathrm{*}}$where $\mu_s$ and $\mu_t$ are, respectively, the Starlink's and terrestrial network's average latency (ms).} \\
\end{tabular}
\end{center}
\end{table*}

%% file: tables/2022-06-05_end_avg_throughput.tex
\begin{table*}[tbp]
\begin{center}
\caption{Average Throughput (Mb/s) across Regions and Protocols.}
\label{table:reg_avg_throughput}


\begin{tabular}{|l|rrrr|rrrr|}
\hline
 & \multicolumn{4}{c}{Terrestrial} & \multicolumn{4}{|c|}{Starlink} \\
\cline{2-9}
& \multicolumn{2}{c}{TCP} & \multicolumn{2}{c}{UDP} & \multicolumn{2}{|c}{TCP} & \multicolumn{2}{c|}{UDP} \\
\cline{2-9}
& Download & Upload & Download & Upload & Download & Upload & Download & Upload \\
\hline
\hline
N. California & 798.89 & 64.21 & 804.97 & 104.42 & 107.77 & 6.26 & 201.42 & 10.41 \\
\hline
Tokyo & 647.55 & 42.69 & 807.79 & 99.96 & 87.28 & 6.24 & 205.72 & 11.18 \\
\hline
London & 644.71 & 43.98 & 806.91 & 103.49 & 78.51 & 6.08 & 203.56 & 10.16 \\
\hline
Sydney & 238.38 & 37.66 & 804.55 & 98.67 & 70.72 & 7.16 & 200.08 & 10.10 \\
\hline
Singapore & 475.39 & 43.25 & 805.45 & 101.02 & 78.11 & 6.33 & 202.50 & 10.56 \\
\hline
Sao Paulo & 481.63 & 47.68 & 807.84 & 103.21 & 73.83 & 6.13 & 205.39 & 10.49 \\
\hline
Bahrain & 398.05 & 54.07 & 810.59 & 103.20 & 74.49 & 6.14 & 194.27 & 9.90 \\
\hline
Mumbai & 346.69 & 51.35 & 803.79 & 101.65 & 66.43 & 6.29 & 196.20 & 10.33 \\
\hline
Cape Town & 318.90 & 55.62 & 816.03 & 105.58 & 68.51 & 6.34 & 193.67 & 10.07 \\
\hline
\end{tabular}

\end{center}
\end{table*}

%% file: sections/experiments/scp.tex
\begin{figure}[t]
    \centering
    \includegraphics[width=0.8\linewidth]{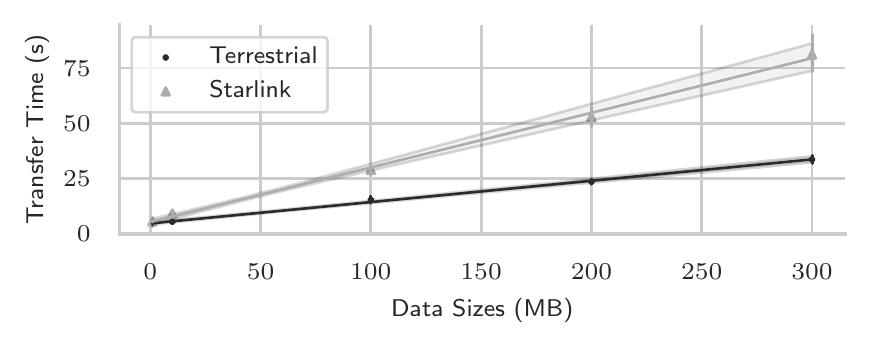}
    \caption{Transfer time as a function of SCP data size (translucent area outlines regression estimates).}
    \label{fig:scp_regress}
\end{figure}

We have also examined bulk data transfer through SCP. As displayed in Fig.~\ref{fig:scp_regress}, compared to the terrestrial network, there is a much stronger upward trend for Starlink in transfer time with higher variance when data sizes grow. While there is almost no difference for the 1 MB data size likely due to SCP overhead, SCP over Starlink almost doubles in transfer time compared to the terrestrial network for the 200 MB data size. This suggests that network dynamics of the satellite hop accumulate overtime and hence have amplified the impact with larger data size in SCP.

%% file: sections/experiments-wild.tex
\section{Starlink in the Wild}
\label{sec:starlink_in_wild}

Serving remote areas has always been a key target of satellite communication systems, especially for Starlink. We have performed a series of measurements in these challenging environments with no terrestrial Internet coverage, not even cellular or power services. We now present the results for two representatives: an estuary in the far north, and a deep valley surrounded by steep mountains.

\subsection{Northern Shoreline}

\begin{figure}[t]
    \centering
    \subfigure[Dish F facing towards a clear sky.]{
        \includegraphics[trim={0 0.5cm 0 0.6cm},clip,width=.4\linewidth]{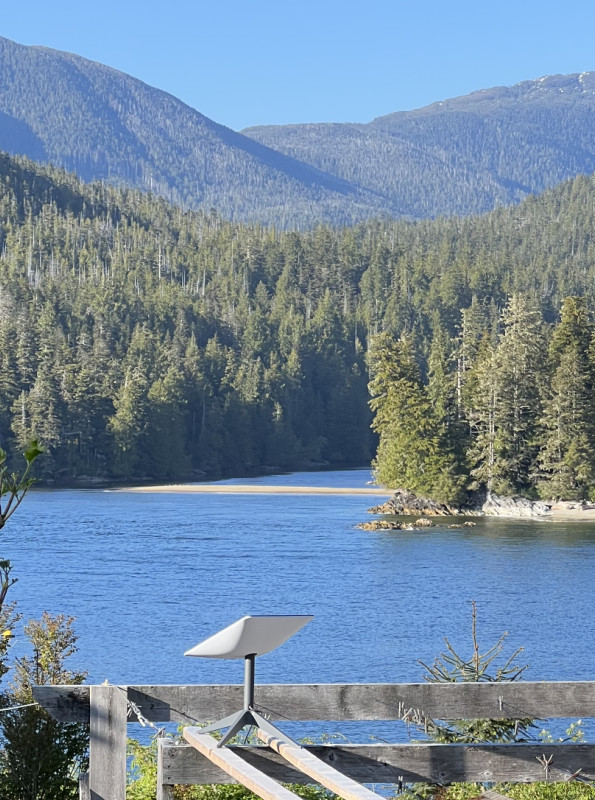}
        \label{fig:koeye_point_dish}
    }
    \subfigure[Google satellite map.]{
        \includegraphics[trim={8.1cm 4cm 8.3cm 1.2cm},clip,width=.4\linewidth]{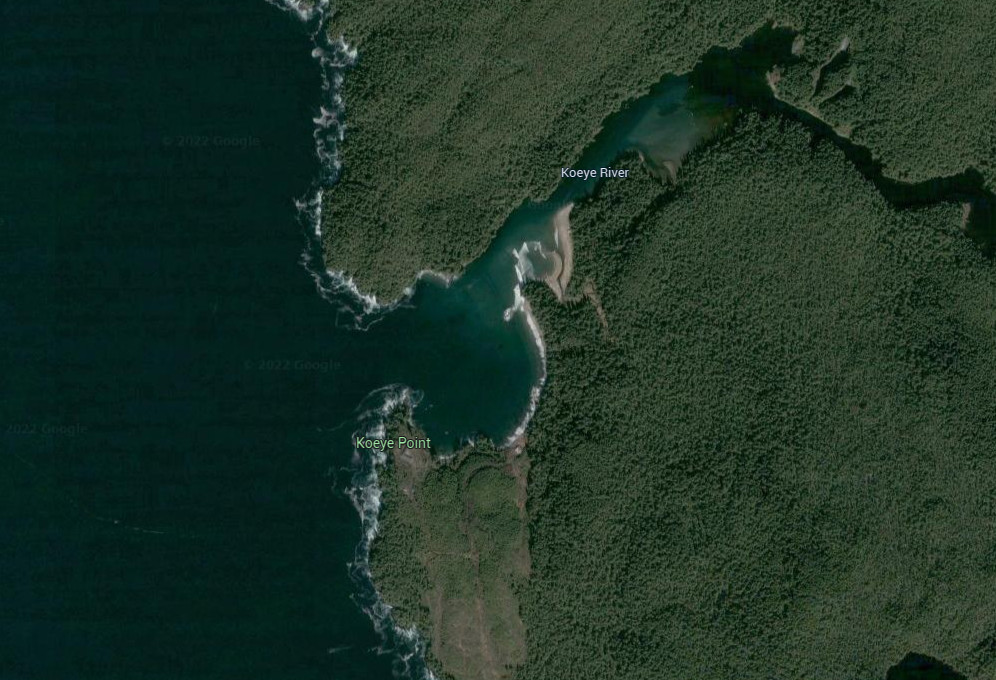}
        \label{fig:koeye_point_google_map}
    }
    \caption{(a) Dish F (Gen 2) setup at (b) Koeye point (Estuary of Koeye River).}
    \label{fig:koeye_point}
\end{figure}

From Vancouver to the estuary (Koeye Point) as seen in Fig.~\ref{fig:koeye_point_google_map} involves a 2-hour flight to Bella Bella in the north and then a 2-hour motor boat ride along the Pacific coastline. Our trip is for a collaborative project with biologists and the Heiltsuk native people for salmon and forest conservation. Building a communication service plays an important role in the project. Our test is based on dish F shown in Fig.~\ref{fig:koeye_point_dish}, a Gen 2 that is easier to carry over on airplanes. 

The simple lodge we were living at the Koeye Point was built by the Heiltsuk people. Near the Pacific shoreline, it has a nice sky clear of obstructions. There is low light pollution, allowing clear sight of satellites (100\%) and space stations orbiting above. Unfortunately, even though this location is recently covered by Starlink with Dish F being usable for simple day-to-day applications, its download throughput is about 68\% lower than that in the urban city and the latency is about 11\% to 30\% longer, with interruptions from time to time. This is likely due to there being fewer satellites and GSes being able to cover the northern areas (Shown in Fig.~\ref{fig:starlink_current_routing}) and the inclination constraint of the first shell. As such, more efforts are required to find the next satellite to handover and set up the GSL and UL.

\subsection{High-Elevation Deep Valley}

\begin{figure}[t]
    \centering
    \subfigure[The surrounding terrain map.]{
        \includegraphics[trim={0 0 0 0},clip,width=.4\linewidth]{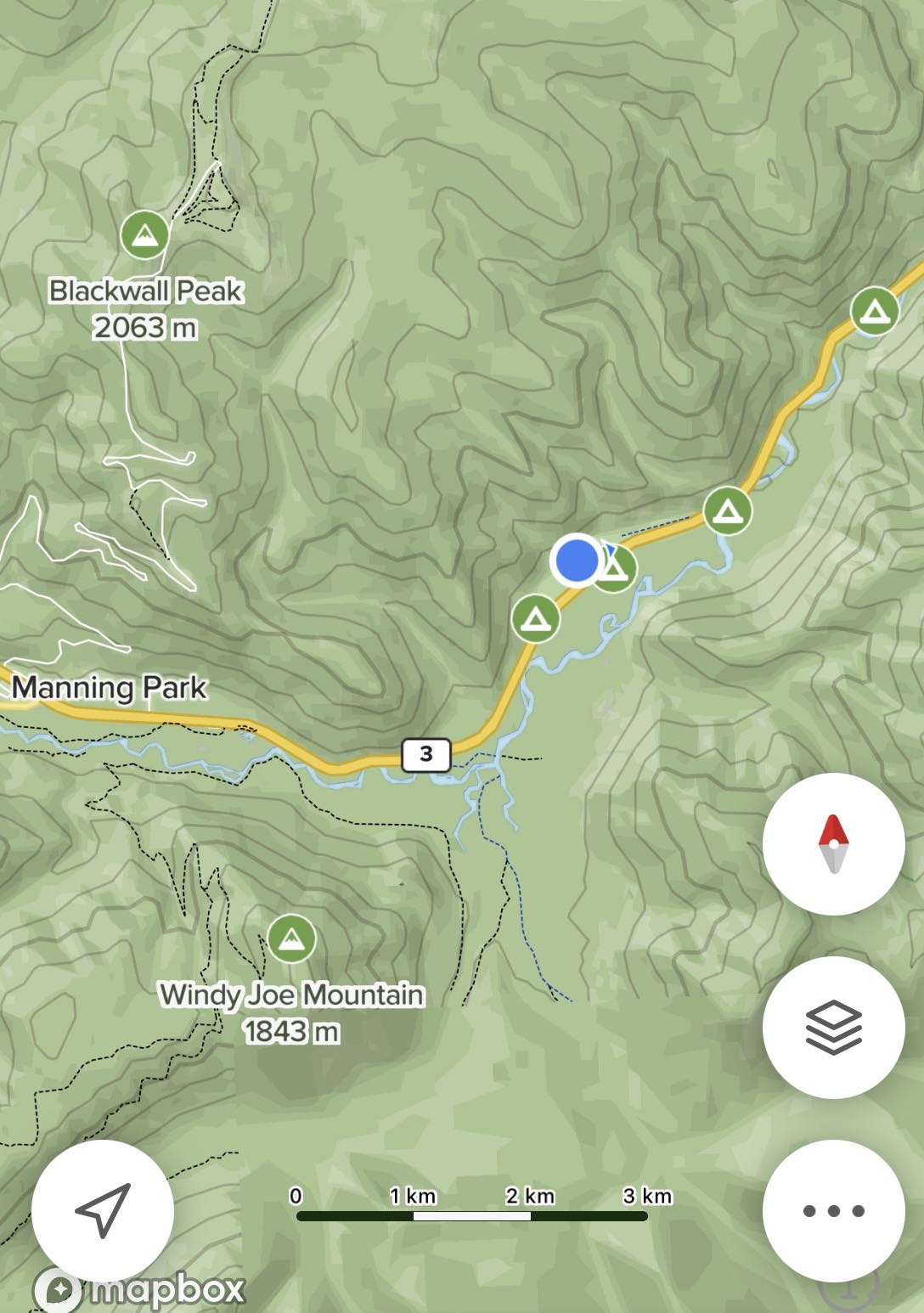}
        \label{fig:manning_park_terrain_map}
    }
    \subfigure[High elevation mountains.]{
        \includegraphics[trim={0 0 0 15.2cm},clip,width=.4\linewidth]{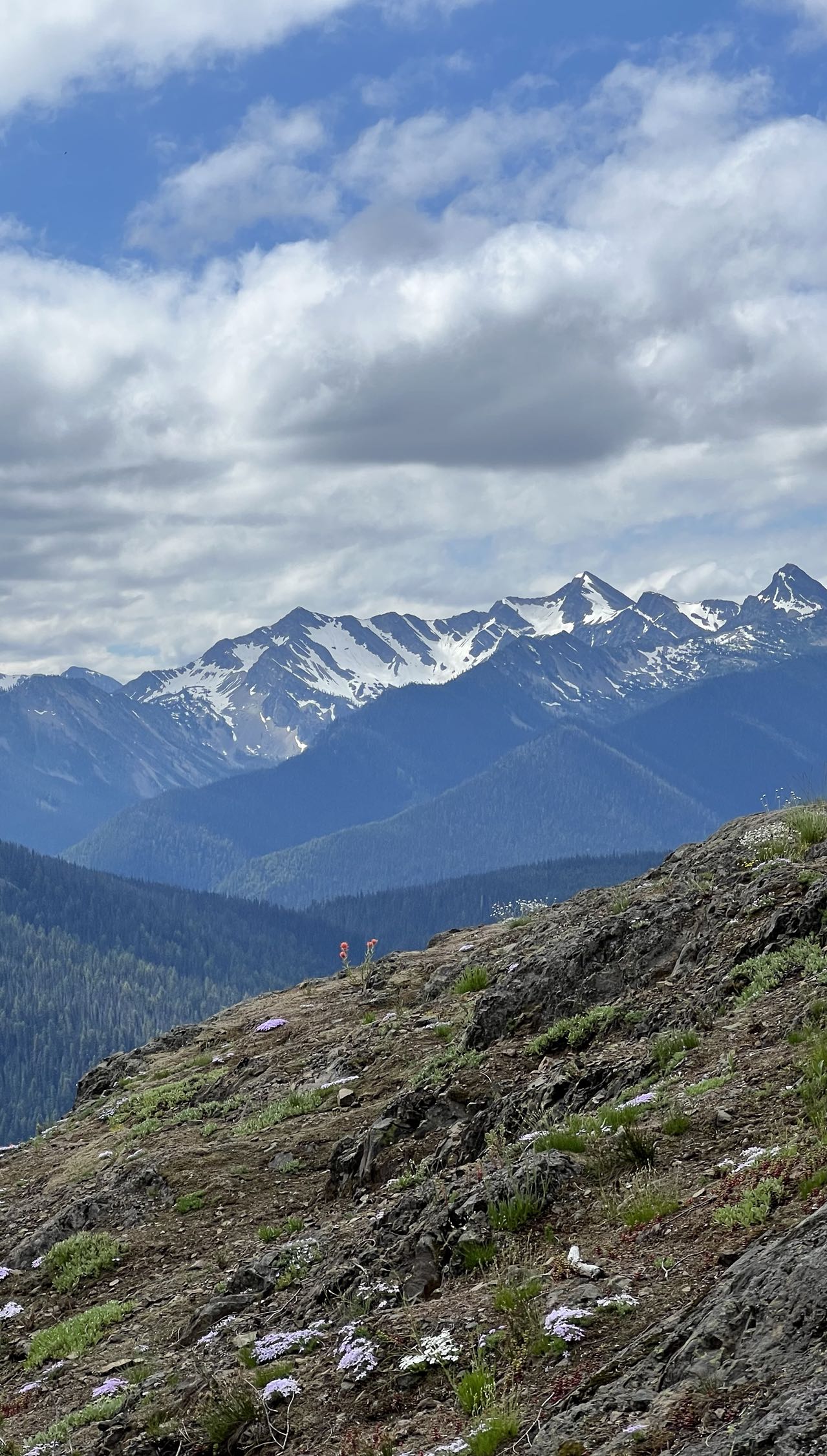}
        \label{fig:manning_park_mountains}
    }
    \caption{The  terrain of the valley in Manning Park. (a) Terrain map; (b) surrounding mountains.}
    \label{fig:manning_park_terrain}
\end{figure}

Our second test is at the center of Manning Park, BC, which has a rough terrain at around 1,000 meters above the sea surrounded by mountains over 2,000 meters as shown in Fig.~\ref{fig:manning_park_terrain}. It is a 3-hour driving distance from Vancouver and has basically no cellular network coverage, not to mention Internet access. We carried over dish C in our car to the site with the portability option turned on, allowing the dish to be used out of its original registered location.

\begin{figure}[t]
    \centering
    \includegraphics[trim={3cm 2.2cm 3cm 5cm},clip,width=.7\linewidth]{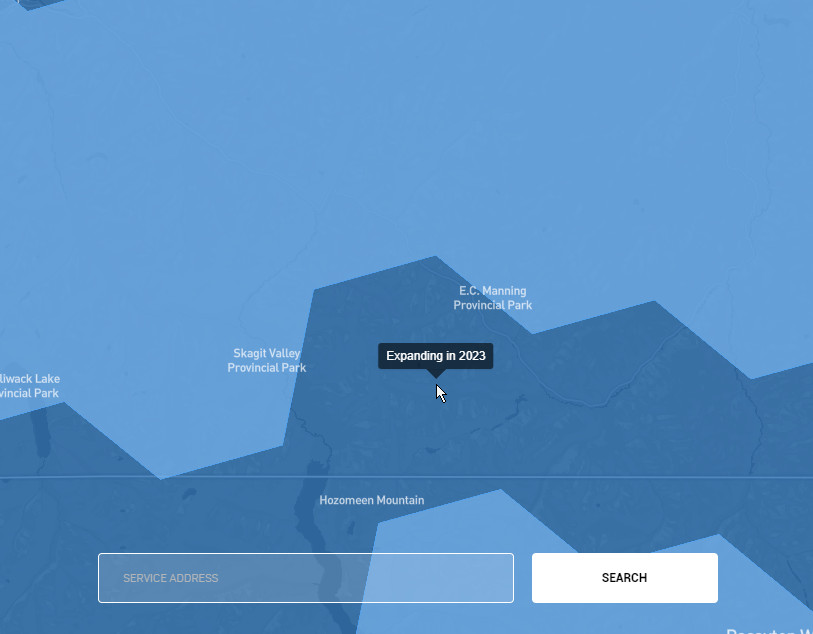}
    \caption{Starlink coverage map around Manning Park.}
    \label{fig:manning_park_starlink_map}
\end{figure}

Despite this valley not yet being officially serviced by Starlink till the end of 2023 as shown in Fig.~\ref{fig:manning_park_starlink_map},\footnote{\url{https://www.starlink.com/map}} we were able to connect to satellites without denials, albeit with poor services. Latency ranged from 90-350 ms on average across the different regions with huge fluctuations over 1,000 ms. Throughput was on average 13 Mb/s and 4 Mb/s with rare bursts to 100 Mb/s and 20 Mb/s for download and upload, respectively. Frequent outages happened every 1-3 minutes, even though the App reports only a 2\% obstruction ratio. Given the terrain, we believe that any LEO satellite, if not being directly above, would be obstructed by the mountains. The clear short-distance visibility to the sky within the valley would mislead the App's algorithm in calculating the obstruction. As a matter of fact, dish C used to be placed within a balcony in the urban city, suffering from a 24.9\% obstruction ratio, but its end user experience there was nearly as good as dishes A and B. This suggests that location (and terrain in particular) plays an important role and not all obstructions are equal to the end user experience. With more satellites being deployed and the valley becoming officially serviceable, the service quality would be improved, if not to the level of flat open areas.

It is also worth noting that both remote locations have not been connected to the power grid, nor would be in the near future. Hence we have to rely on a solar-diesel hybrid power system at the estuary and a battery power pack in the valley. Given the current Starlink kit's power consumption, a typical 1,500 Ah power pack  lasts around only 1 hour.

%% file: sections/future.tex
\subsection{Stability of Bent-Pipe}

\input{sections/experiments/web-streaming}

\subsection{Mobility Potentials}

\begin{figure}[t]
    \centering
    \includegraphics[width=.7\linewidth]{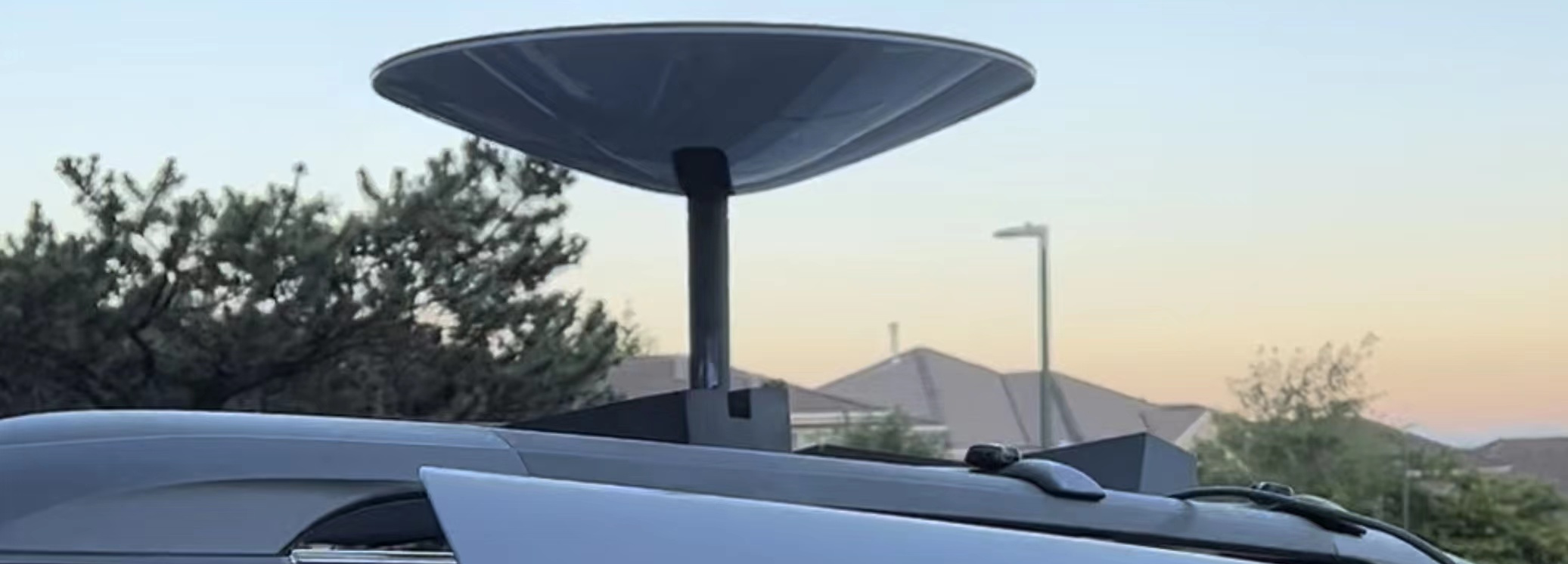}
    \caption{A Gen-1 dish secured on the roof of a minivan.}
    \label{fig:dish_van}
\end{figure}

\begin{figure}[t]
    \centering
    \includegraphics[width=.8\linewidth]{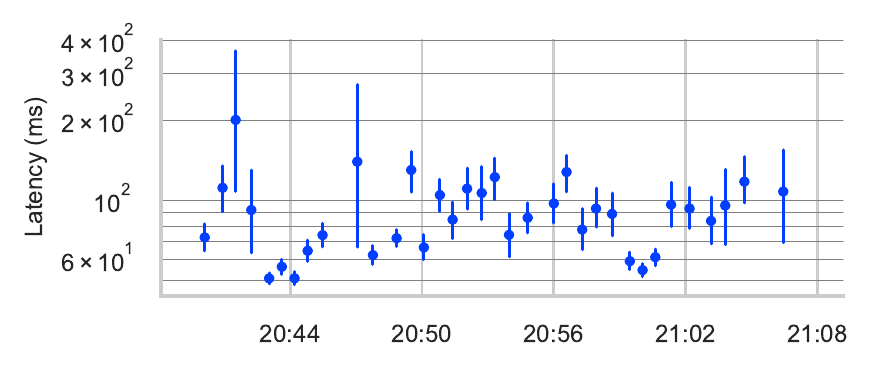}
    \vspace{-2mm}
    \caption{Latency to N. California while in motion (vertical lines correspond to 95\% confidence intervals).}
    \label{fig:mobility_ping}
\end{figure}

\begin{figure}[t]
    \centering
     \includegraphics[width=.8\linewidth]{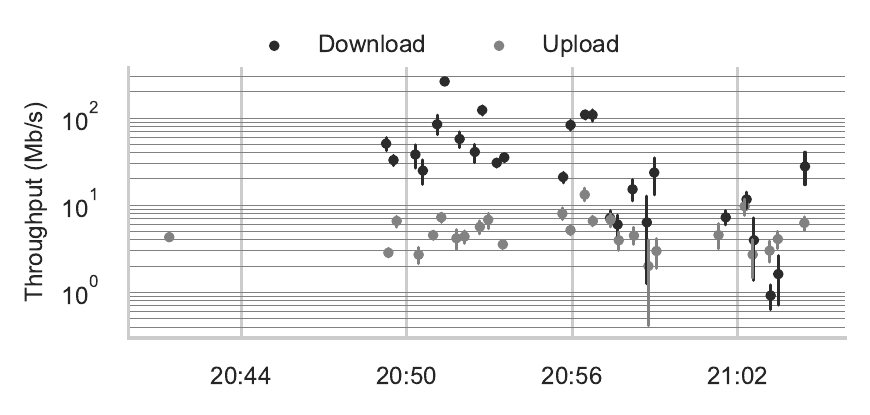}
    \vspace{-2mm}
    \caption{Throughput to N. California while in motion (vertical lines correspond to 95\% confidence intervals).}
    \label{fig:mobility_iperf}
\end{figure}

FCC has recently authorized Starlink Internet for use on vehicles in motion,\footnote{\url{https://www.cnbc.com/2022/06/30/fcc-approves-spacex-starlink-service-to-vehicles-boats-planes.html}} though it has yet to be fully supported by Starlink. Hence, we have run a short test at this stage to gauge the work needed to get Starlink ready for full in-motion connectivity.\footnote{This test was for academic research only. Currently, Starlink does not suggest general use of its dishes in motion, although there have been trials, notably in Ukraine, and Starlink also plans to release an RV version.} We secured dish C with two crossbars on the roof-rack of a mini van (see  Fig.~\ref{fig:dish_van}), with the dish's portability option being turned on. In our 30-minute test, the car went at 40 to 70 km/h around three different driving routes with twists and turns and then a full stop near the end. Most of the time, the dish was flat towards the sky with some minor self-adjustment. Note that the dish in the stationary setup is mostly tilted towards one direction. Hence, likely the dish can sense motion and accordingly adjust the flat position, which works best for omnidirectional reception. 

The current Starlink dish works while in motion but still has a long way to go to match the current stationary setup. We have observed outages at the median every 16.5 seconds for 5 seconds, and some every 162 seconds lasting as high as 36 seconds (95-th percentile). Latency as shown in Fig.~\ref{fig:mobility_ping} is on average 100 ms with the highest being 2,800 ms at 20:42. The latency spikes greater than 200 ms at least twice every minute on average with as high as 8 times in a minute. This would frequently interrupt applications that require sub-200 ms for satisfying experiences. Considering that a stationary dish averages around 50 ms as we have discussed earlier, the latency while mobile is high with larger fluctuations. Some of these latency increases could be due to sudden orientation changes as the car turns around corners or intermittent obstructions from trees passing by. For example, a right-hand turn occurred at around 20:40, which coincides with the sharp spike in latency at around the same time. Download throughput has reached similar averages to the stationary test at around 80 to 100 Mb/s but immediately drops off near the end, as shown in Fig.~\ref{fig:mobility_iperf}. Upload throughput, interestingly, has been able to keep up around 5 to 10 Mb/s, on par with that of the stationary setup. Both observe a noticeable drop at around 20:58, likely due to a disconnection from a satellite. The last 6 minutes have been mostly a straight drive in a valley, suggesting that perhaps the dish was forced to switch satellites faster than expected. 

Note that the Starlink kit typically takes 3-7 minutes to boot up and be Internet ready, but we have observed cases over 20 minutes. Such a long time will be a barrier for mobile use.

\subsection{Global Coverage? Technical and Cultural Implications}

To achieve truly global coverage, more Starlink satellites are needed, including other shells covering the North and South Poles. Our remote area experiments were  done 14\textdegree{} south of the Arctic Circle (66\textdegree{}34'N), which is already almost out of the current service area in North America. Inter-satellite links are necessary to be enabled, too, for efficient cross-continent communications and for robust operation when disasters happen near the Earth's  surface. 

\begin{figure}[t]
    \centering
    \subfigure[Aerial view with 30m tall trees]{
        \includegraphics[width=.7\linewidth]{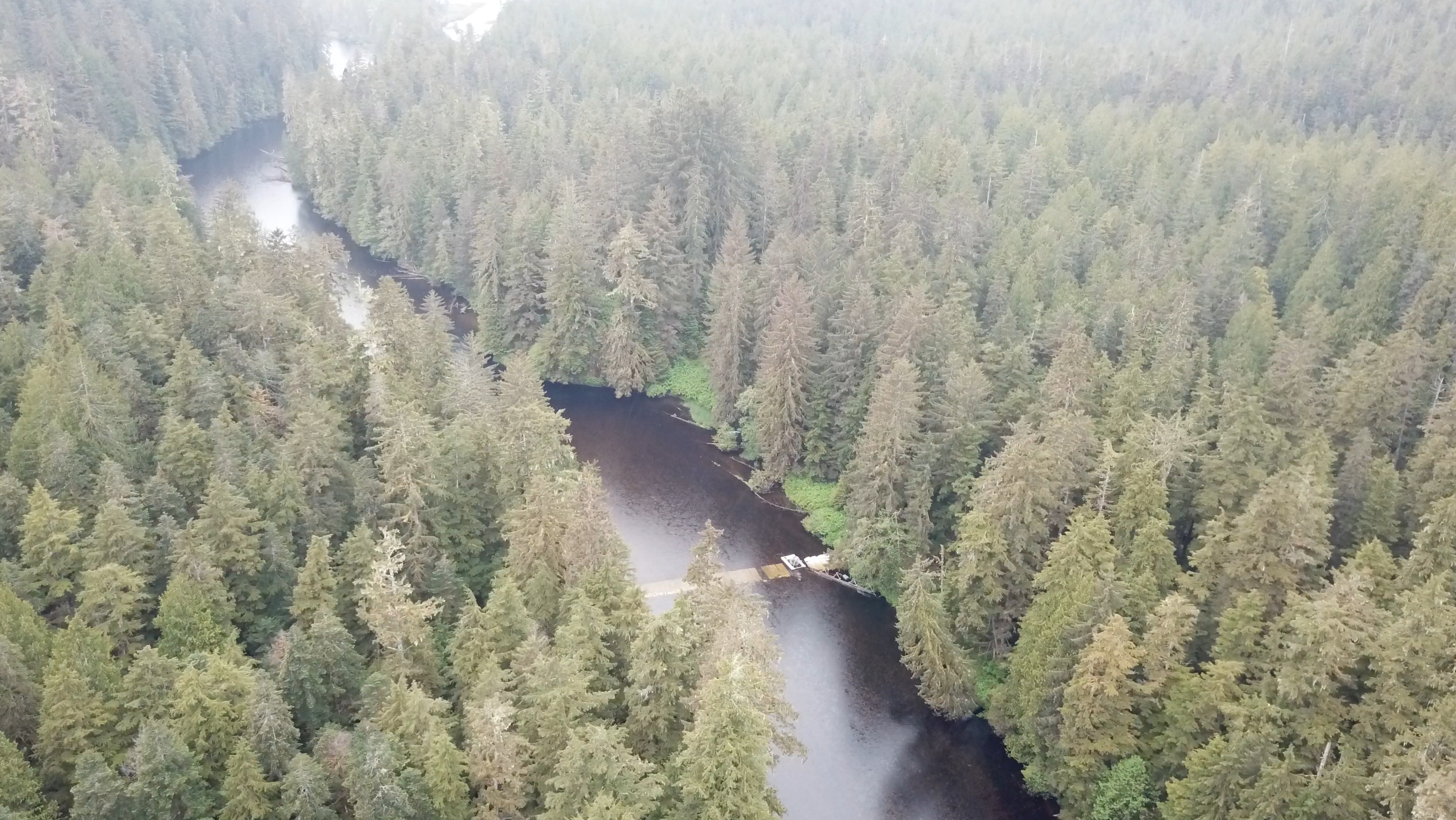}
        \label{fig:koeye_weir_trees}
    }
    \subfigure[Weir setup and solar panels]{
        \includegraphics[trim={3cm 3cm 3cm 3cm},clip,width=.7\linewidth]{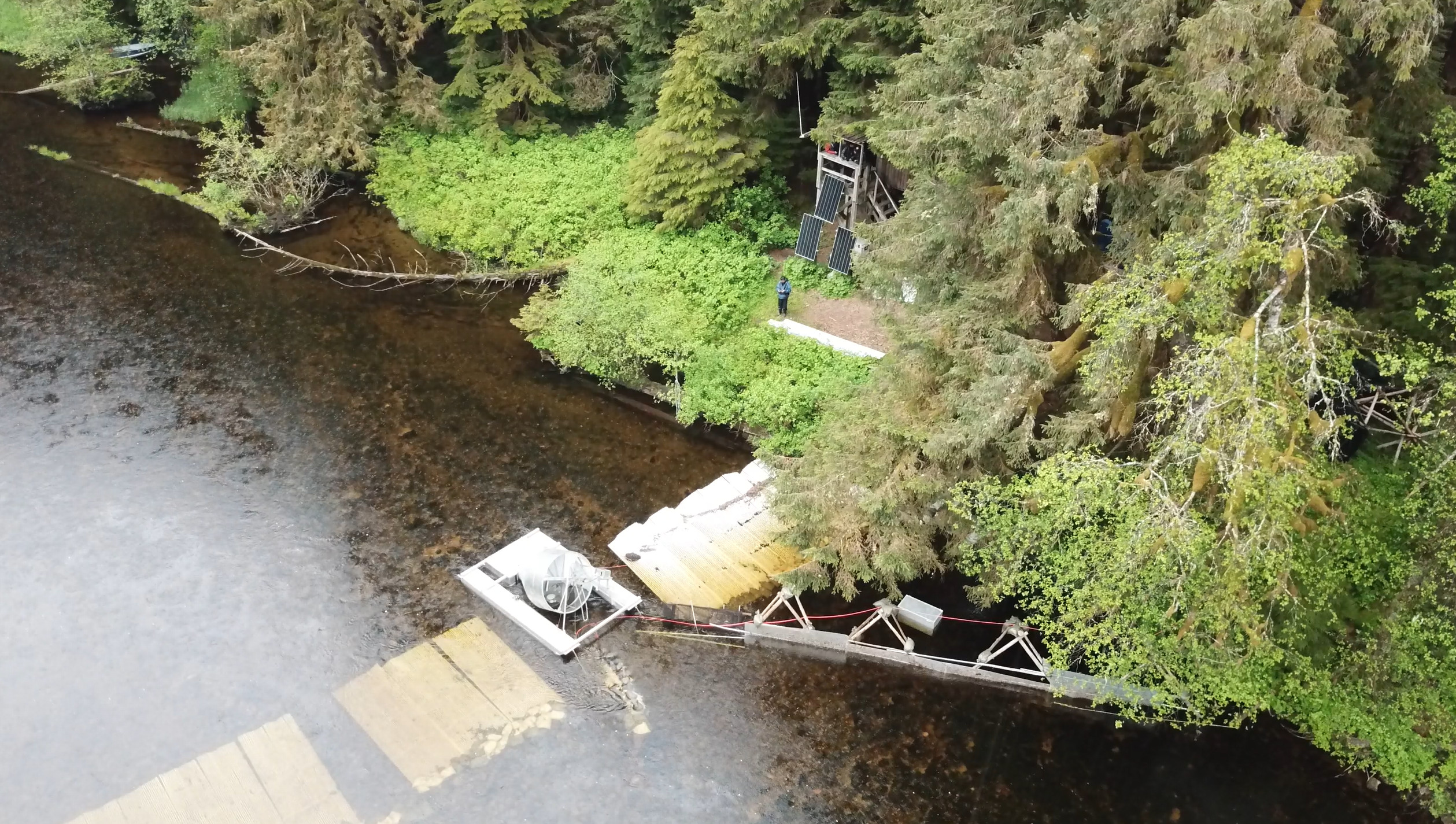}
        \label{fig:koeye_weir_setup}
    }
    \caption{An aerial view of the weir site upstream of the Koeye River in the Great Bear rainforest.}
    \label{fig:koeye_weir}
\end{figure}

Even if these technical issues would eventually be overcome, we have realized many challenges during our site visit deep into the Great Bear temperate rainforest at the estuary's upstream, where we planned to install a Starlink dish  near the Heiltsuk  aboriginal people's reconstructed weir for a fish trap and its monitoring system (Shown in Fig.~\ref{fig:koeye_weir_setup}).

Practical challenges include clearance, wildlife, and power. The old-growth trees are 30 meters tall as can be seen in Fig.~\ref{fig:koeye_weir_trees}, and the only clear sky available is at the weir on the riverside, which is subject to flooding. Wildlife such as bears, wolves, and birds could  easily cross the narrow weir and disrupt the dish. As a matter of fact, one hundred distinct individual grizzly bears have been identified in the Koeye watershed. Also, solar is the only source of power available in such a remote area in the deep forest (The diesel generator is not allowed nor practical). As such, intermittent operations are inevitable as the dish needs a max of 145 watts of power if it does not enter snow melting mode.

Cultural challenges are more complicated, which we have not considered before the site visit.  The Koeye river and its surrounded temperate rainforest is a holy land of the Heiltsuk native people, hosting their traditional {\em big house}. The people there bring their set of conflicts with their traditional values. Signal pollution includes more philosophical thought of pristine wilderness and as more of the world is connected, fewer  places will exist as a getaway from constant connectivity. Light pollution is commonly known nowadays, and a global coverage brought by the  LEO satellite constellation would bring another pollution of invisible electromagnetic waves, the {\em WiFi pollution}. This is not in terms of the technological aspect but in the social aspect. This has been the major concern of the biologists working with us. As a matter of fact, in the true wild, individuals will have to always stay alert and mostly take care of themselves; hence, the distraction of using mobile apps would overshadow the benefit of staying connected. Policies, professional guidelines, and ethics across different disciplines and cultures are therefore urgently needed for the forthcoming global Internet coverage on Earth and in space.

%% file: sections/experiments/web-streaming.tex
\begin{figure}[t]
\centering
\includegraphics[width=0.9\linewidth]{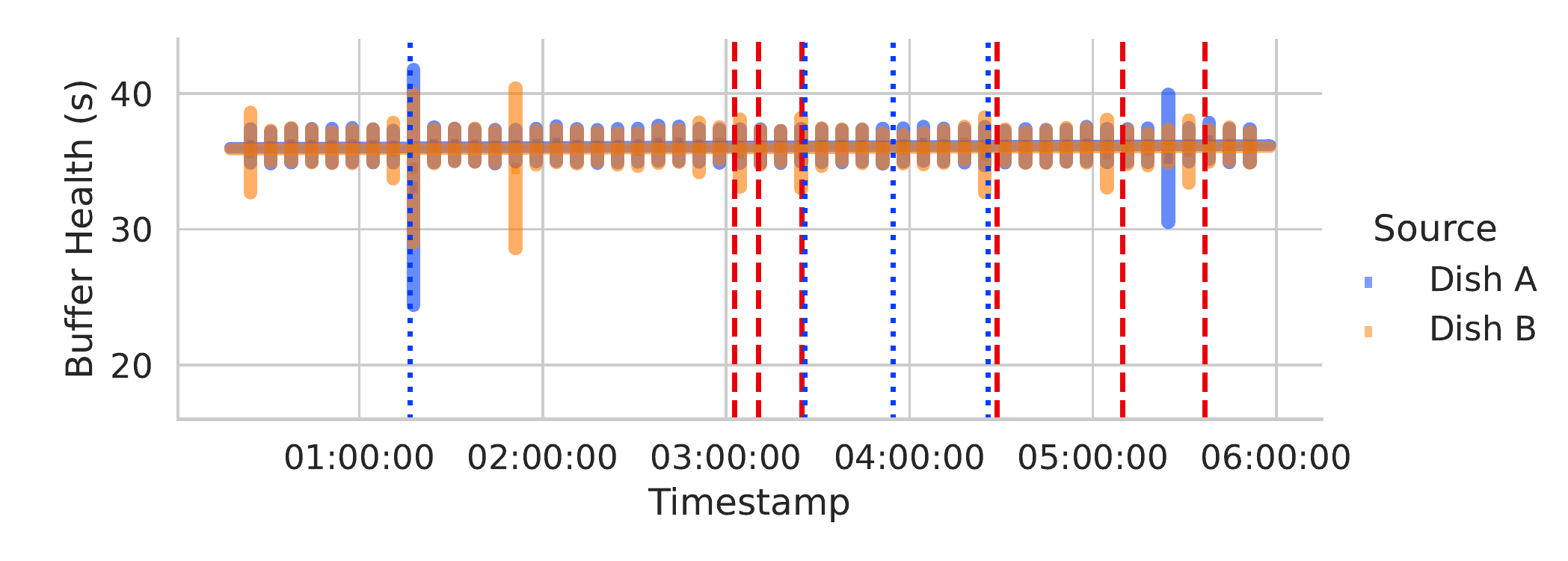}
\caption{Buffer health over time for 2-dish synchronous streaming. Each bar outlines the standard deviation for 7-min samples. Outages are highlighted by vertical dotted lines.}
\label{fig:0616_zhy_sami_bh}
\end{figure}

Our early experiments have suggested that, despite being reasonably good in throughput, the bent-pipe is often unstable in communication. To closely examine its stability, we have done a stress test under continuous streaming of an ultra-high-resolution 8K (7,680 x 4,320 pixels) YouTube video.\footnote{\url{https://www.youtube.com/watch?v=KcMlPl9jArM}} We place two dishes (A and B) in nearby locations within the coverage of the same satellite. In other words, they share the same GSL but have their respective ULs. We then perform YouTube streaming synchronously to them, each being attached to a PC to playback the video. The synchronous streaming sessions last over 24 hours by loop playing back the video, and the PCs' browser buffers were cleaned up after each playback to ensure continuous data downloading. We use a script to automatically scrape YouTube's built-in tool \texttt{Stats for nerds} for real-time data statistics \cite{uran_starlink_2021}, including connection speed, network activity, and buffer health. The buffer health here represents the length (in seconds) of pre-loaded video content, which, being sensitive to network outages, is directly correlated with the viewer's perceived quality.

\renewcommand{\arraystretch}{1.1}
\begin{table}[t]
\begin{center}
\caption{Outage counts for 2-dish synchronous streaming.}
\label{table:sync_web_stremaing_summary}
\begin{tabular}{c|c|c|c|c}
 \hline
  & 
 \begin{tabular}[c]{@{}l@{}}
 Buffer\\Outage
 \end{tabular} &
 \begin{tabular}[c]{@{}l@{}}
 App Reported\\Outage
 \end{tabular} & 
 \begin{tabular}[c]{@{}l@{}}
 Overlapped\\Outage
 \end{tabular} & 
 \begin{tabular}[c]{@{}l@{}}
 Overlapped\\Outage Ratio
 \end{tabular}
 \\ 
 \hline
 Dish A & 8 & 8 & 3 & 37.50\%\\  
 Dish B & 19 & 13 & 3 & 15.79\%\\
 \hline
\end{tabular}
\end{center}
\end{table}

As can be seen from Fig.~\ref{fig:0616_zhy_sami_bh}, both dishes have experienced a series of buffer outages in this synchronous test, some of which are overlapped. For instance, two pairs of buffer outages occurred at 01:18 and 03:26 simultaneously for both dishes, and their Starlink Apps both reported network outages at the same time. The overlapped outages indicate that there can be a common factor, the bent-pipe, causing them between the two dishes. Table.~\ref{table:sync_web_stremaing_summary} summarizes the outage events, from which we can confirm that a non-negligible amount of outage events are caused by their bent-pipes (over 37.5\% for A and 15.79\% for B). The data also show that Dish A is more stable in streaming than B, which is likely because it has clearer visibility. There are more trees around dish B, so the Starlink App has reported a 2.7\% and a 4.7\% obstruction ratio for Dish A and B, respectively. Hence, even in the same service area, different Starlink users may experience different UL quality and hence network service quality.

%% file: sections/conclusion.tex
This paper has presented our initial measurements on the network characteristics of today's Starlink from an end-user's perspective. Many of the issues identified from our study are not just confined to Starlink, but are common for LSNs. Our observations therefore will help with optimizing the LSN deployment and operation, and will also help developers and users customize their networks and applications. 

LSN services are rapidly evolving, so we will be continuing to monitor the performance evolution of Starlink's Internet service. For instance, 
the weather or cloud effects on Starlink can be further expanded with rain probability \cite{noauthor_recommendation_nodate} and cloud attenuation\cite{yuan_high-resolution_2019} models in future experiments. They can include optimization methods that take into account the probability of weather attenuation and act accordingly to provide uninterruptible service. For the bent-pipe's stability, a future repeat experiment could be conducted by mounting the dishes to the roof of each respective house to mitigate the obstruction ratio variable. 

We are also interested in the effectiveness of the upcoming new functions, e.g., satellite-to-satellite links, multiple bent-pipe, and mobility support, to name but a few. We are also interested in the seamless integration of LSN and terrestrial Internet and 5G cellular networks towards truly global coverage and the related inter-disciplinary issues.